\documentclass[aps,prx,twocolumn,superscriptaddress,nofootinbib,notitlepage,longbibliography]{revtex4-2}

\usepackage{amsmath,hyperref,dcolumn,graphicx,color,bm,extpfeil,indentfirst,nicematrix,bbm,centernot,cancel,mathrsfs,comment,ulem,mathtools}
\usepackage{tocbasic}

\DeclarePairedDelimiter\ket{\lvert}{\rangle}
\DeclarePairedDelimiter\bra{\langle}{\lvert}

\newcommand*\diff{\mathop{}\!\mathrm{d}}

\begin{document}
	
	% --- Main text (two columns) ---
	%\documentclass[aps,prx,twocolumn,superscriptaddress,nofootinbib,notitlepage,longbibliography]{revtex4-2}
%\usepackage{graphicx}% Include figure files
%\usepackage{dcolumn}% Align table columns on decimal point
%\usepackage{graphicx,color} % Required for inserting images
%\usepackage{amsmath}
%\usepackage{bm}
%\usepackage{extpfeil}
%\usepackage{indentfirst}
%\usepackage{nicematrix}
%\usepackage{bbm} % to draw empty letter/number
%\usepackage{centernot} % to negate a symbol
%\usepackage[thicklines]{cancel} % to cancel terms in euqation
%\usepackage{mathrsfs} % to use \mathscr
%\usepackage{comment} % to comment paragraph
%\usepackage[normalem]{ulem}
%\DeclarePairedDelimiter\ket{\lvert}{\rangle}
%\DeclarePairedDelimiter\bra{\langle}{\lvert}
%\DeclarePairedDelimiter\braket{\langle}{\rangle}
%\newcommand*\diff{\mathop{}\!\mathrm{d}} % integral d
%\newcommand*\Diff[1]{\mathop{}\!\mathrm{d^#1}} % integral d^2
%\usepackage{mathtools} % For $:=$
%\newcommand{\JX}[1]{\textcolor{red}{#1}}
%\newcommand{\XL}[1]{\textcolor{blue}{#1}}
%
%\begin{document}

\title{Non-Abelian Quantum Metric Governed Topological Boundary-Mode Localization}

\author{Xing-Lei MA}
\author{Jin-Xin Hu}\thanks{jhuphy@ust.hk}
\author{K. T. Law}\thanks{phlaw@ust.hk}

\affiliation{Department of Physics, Hong Kong University of Science and Technology, Clear Water Bay, Hong Kong, China}

\date{\today}
\begin{abstract}
    
    The presence of localized boundary modes is an unambiguous hallmark of topological quantum matter. While these modes are typically protected by topological invariants such as the Chern number, here we demonstrate that the {\it quantum metric length} (QML), a quantity inherent in multi-band topological systems, governs the spatial extent of flat-band topological boundary modes. We introduce a framework for constructing topological flat bands from degenerate manifolds with large non-Abelian quantum metric and find that the boundary modes exhibit two sequential phases of spatial behaviors: a conventional oscillatory decay arising from bare band dispersion, followed by another exponential decay controlled by quantum geometry. Crucially, the QML, derived from the non-Abelian quantum metric of the degenerate manifolds, sets a lower bound on the spatial spread of boundary states in the flat-band limit. Applying our framework to concrete models, we validate the universal role of the QML in shaping the long-range behavior of topological boundary modes. Furthermore, by tuning the QML, we unveil extraordinary non-local transport phenomena, including QML-shaped quantum Hall plateaus and anomalous Fraunhofer patterns. Our theoretical framework paves the way for engineering boundary-mode localization in topological flat-band systems.
        
\end{abstract}

\maketitle

%	\tableofcontents

\section{\label{sec_1} Introduction}
%\emph{Flat band physics with non-trivial quantum geometry.}

Geometry endows electronic states with new characteristic scales \cite{PhysRevLett.82.370,PhysRevB.56.12847,PhysRevB.62.1666,PhysRevLett.122.166602,onishi2024fundamental,Verma2026}, fundamentally reshaping conventional paradigms—particularly in the flat-band regime~\cite{bistritzer2011moire,cao2018correlated,PhysRevLett.121.266401}. With their highly quenched kinetic energy, flat bands herald a paradigmatic shift, as traditional frameworks for understanding crystalline properties break down when the Fermi velocity vanishes. Consequently, the geometric properties of Bloch states have profound implications, ranging from unconventional superfluidity~\cite{peotta2015superfluidity,PhysRevLett.117.045303,PhysRevLett.123.237002,torma2022superconductivity,julku2021quantum,PhysRevLett.132.026002,xie2020topology,jiang2025superfluid} and optical spectral weight~\cite{verma2021optical,mao2023diamagnetic} to density wave instabilities~\cite{han2024quantum,jiang2023pair,hofmann2023superconductivity}.

At the heart of these phenomena lies the quantum geometric tensor, $\mathcal{Q} \equiv \mathcal{G} + \frac{1}{2i}\Omega$, where the real part $\mathcal{G}$ (quantum metric) and the imaginary part $\Omega$ (Berry curvature) encodes distinct geometric properties of Bloch states~\cite{resta2011insulating,PhysRevLett.131.240001,liu2025quantum,yu2024quantum}. While Berry curvature characterizes the phase difference that determines the global band topology~\cite{RevModPhys.82.1959}, the quantum metric defines the infinitesimal distance between states in the parameter space \cite{provost1980riemannian,guo2024majoranazeromodesliebkitaev}. In the past several decades, the interplay between these geometric counterparts has raised wide interest and been explored in several dimensions: ideal quantum geometry in fractional Chern insulators~\cite{PhysRevLett.127.246403,shavit2024quantum}, inequality relations between them~\cite{PhysRevB.104.045103,kruchkov2022quantum,herzog2022superfluid}, and their manifestation in optical responses~\cite{onishi2025quantumweightfundamentalproperty,chen2022measurement}, etc. In the context of the bulk-boundary correspondence~\cite{PhysRevLett.71.3697}, it has been firmly established that localized boundary modes are protected by bulk topological invariants since the discovery of the integer quantum Hall effect~\cite{PhysRevLett.45.494}.
%In particular, \XL{'Not clear how this paragraph connects to the previous one. What are you particularizing here?'} since the discovery of the integer quantum Hall effect~\cite{PhysRevLett.45.494}, it has been firmly established that localized boundary modes are protected by bulk topological invariants—a fundamental principle known as the bulk-boundary correspondence~\cite{PhysRevLett.71.3697}. 
Therefore, a fundamental question arises: Does the bulk-states quantum metric, which generally hosts a non-Abelian structure, play any role in the localized boundary modes? 

%Recently, in a separate work~\cite{guo2024majoranazeromodesliebkitaev}, we pointed out that in flat-band topological superconductors, the long-range behavior of Majorana Zero Modes in the Lieb-Kitaev model is governed by the quantum metric length (QML). However, a comprehensive study inspecting the origin, universality, and experimental consequences of QML in generic topological systems remains unexplored.

In this work, we develop a framework to construct topological flat bands with nontrivial quantum metric, and identify how the quantum metric influences the localization behaviors of topological boundary modes (TBMs). It is found that multi-band TBMs typically exhibit two sequential phases of behaviors: an initial oscillatory decay driven by the band dispersion, followed by an exponential decay arising from the Wannier orbitals encoding the quantum geometry. They are thus dubbed the conventional and the geometric behavior, respectively. It is further proved that in the flat-band limit where the conventional behavior is suppressed, the TBMs spread is lower bounded by the {\it quantum metric length} (QML), which originates from degenerate flat band manifolds with non-Abelian quantum metric. This generalized QML is validated across various flat-band models.

%which is derived from the non-Abelian quantum metric of the degenerate flat bands upon which the topological bands are built.  This generalized QML is validated across various flat-band models. %, where both behaviors persist and QML serves a robust lower bound.

%%%%%%%%%%%%%%%%%%%%%%%%%%%%%%%%%%%%%	
\begin{figure*}
    \centering
    \includegraphics[width=0.9\linewidth]{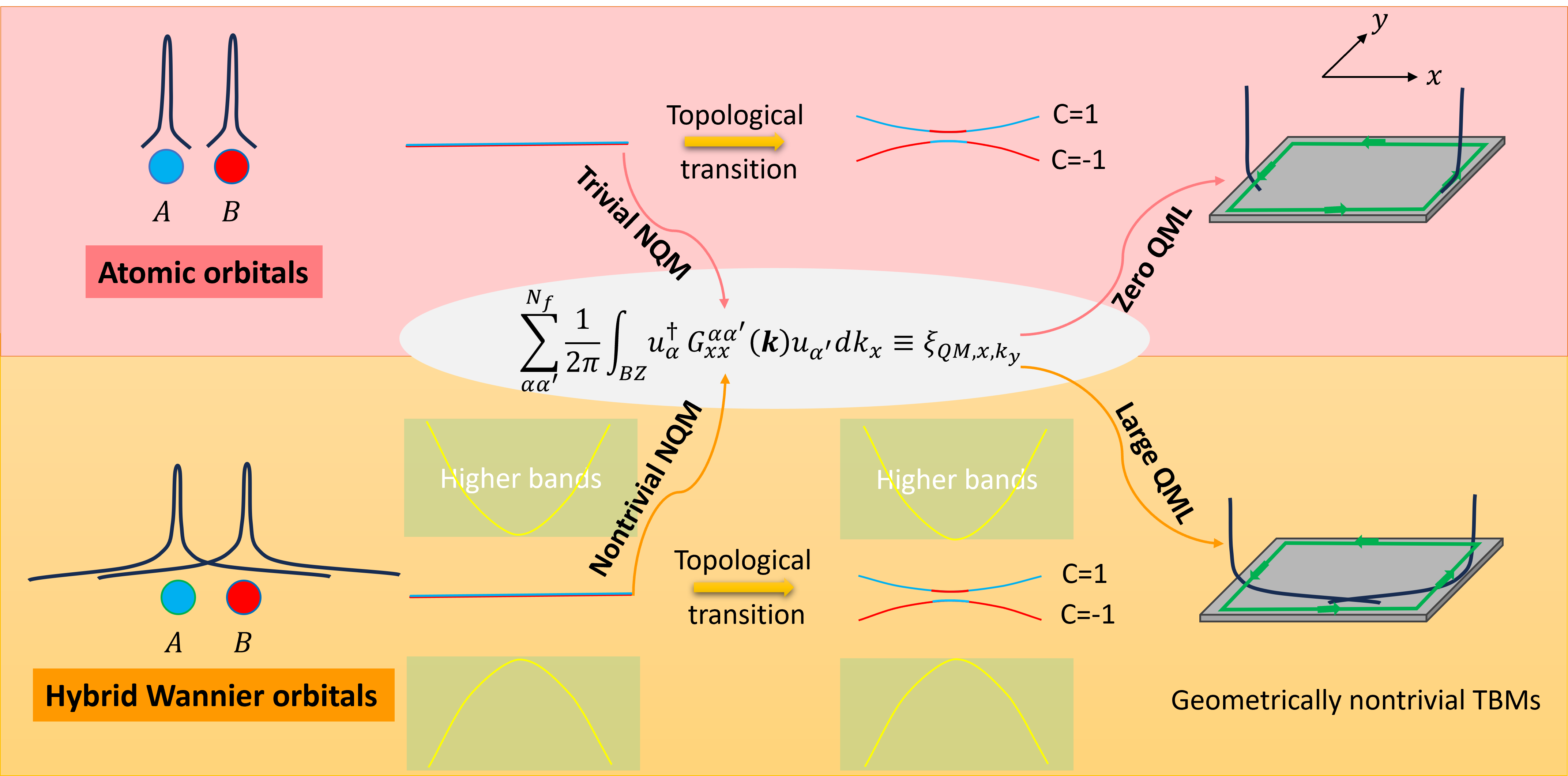}
    \caption{\textbf{An illustration of two paradigms in constructing 2D topological flat bands.} 
    %\textbf{An illustration of constructing 2D flat-band TIs, and the relation between the quantum metric of the degenerate flat bands and the long-range localization of the flat band TBMs.} 
    Upper panel: Decoupled atomic orbitals form degenerate flat bands with trivial non-Abelian quantum metric (NQM), resulting in well-localized TBMs with zero QML after topological transition. Lower panel: Degenerate flat bands with nontrivial NQM, the same process leads to geometrically nontrivial TBMs with long-range localization controlled by the large QML.
    %long-range localized TBMs, which are controlled by the large QML.
    %quantum geometrically nontrivial TBMs with long-range localization controlled by the large QML.
    %, $\xi_{\text{QM},x,k_y}$, in a direction and channel-dependent manner.
}
\label{Fig_Illustration_QML_TBM}
\end{figure*}
%%%%%%%%%%%%%%%%%%%%%%%%%%%%%%%%%

Our framework begins with degenerate flat bands from a parent Hamiltonian $H_{\text{f}}$, and perturbative couplings $H_{\text{c}}$ are introduced to make the flat bands topological. The total Hamiltonian $H_{\mathrm{t}}$ in general takes the form
\begin{equation}
    H_{\mathrm{t}}(\bm{k}) = H_{\text{f}}(\bm{k}) + H_{\text{c}}(\bm{k}),
    \label{Eq_H_tf}
\end{equation}
where the coupling term $H_{\text{c}}$ has subdominant energy scales compared to the original flat-band model. The topological bands inherit the quantum geometry from the original flat bands of $H_{\text{f}}$, profoundly shaping the characteristics of their TBMs.

To be specific, Fig.~\ref{Fig_Illustration_QML_TBM} illustrates the construction of topological bands within two distinct paradigms: (1) In the upper panel, we consider two atomic orbitals (A and B) with identical on-site energies, which form degenerate flat bands with trivial non-Abelian quantum metric. When these orbitals (and consequently the flat bands) are coupled in a 2D array, the degeneracy is lifted and dispersion emerges. During this process, the band gap may close and reopen, leading to an inversion of orbital compositions that yields nontrivial topology, such as topological insulators (TIs)~\cite{PhysRevLett.61.2015}. In this scenario, while the bulk states possess nontrivial topology, we nevertheless regard their gapless boundary modes as geometrically trivial, since their localization is entirely determined by band dispersion. In the flat-band limit, the localization length approaches zero. (2) In the lower panel, we show that, in contrast, when starting from degenerate flat bands with nontrivial non-Abelian quantum metric, the same coupling process produces topological bands whose boundary modes inherit an additional length scale from the spatially extended hybrid Wannier orbitals. Notably, despite of Wannier obstruction on the topological bands~\cite{thouless1984wannier,PhysRevB.74.235111,soluyanov2011wannier,po2018fragile}, the TBMs are safe from it since they are built from the parent manifold that is topologically trivial. Specifically, we demonstrate that this length scale is, in general, controlled by the QML:
\begin{equation}
        \xi_{QM,i, \tilde{\bm{k}}} \equiv  \sum^{N_f}_{\alpha \alpha'} \frac{1}{2\pi}  \int_{\text{BZ}}  {u}_{\alpha}^{*} \mathcal{G}^{\alpha\alpha'}_{ii}(\bm{k}) {u}_{\alpha'} \diff k_i,
    \label{Eq_QML_definition}
\end{equation}
which sets a lower bound for the spatial spread of the TBMs at channel $\tilde{k}$ in the $i$-direction. Here, $\mathcal{G}^{\alpha\alpha'}_{ii}(\bm{k})$ is the non-Abelian quantum metric of the $N_f$ degenerate flat bands~\cite{PhysRevB.81.245129}, and $|u\rangle=(u_1,...,u_{N_f})$ is a vector parameterizing the low-energy effective Hamiltonian. The QML governs the long-range behavior of flat-band TBMs and can readily exceed lattice scales. This principle can be extended to other topological classes, motivating our general framework for engineering TBMs in topological flat bands and for exploring the rich physical consequences enabled by the QML.

\section{Results}
	
\subsection{\label{Sec_TFB_Construction}Construction of topological flat bands with non-trivial quantum metric}

%We present a feasible framework to construct topological flat bands with tunable quantum metric, where the resultant QML can significantly exceed the lattice length scale. The framework can be implemented in two steps. First, we prepare degenerate, exactly flat bands and introduce coupling among them to achieve nontrivial topology. Next, we project the full model onto the flat-band subspace to obtain an effective model. By imposing specific conditions on the original Hamiltonian to locally constrain the flat-band eigenvectors, the low-energy physics manifests clearly, accompanied with nontrivial quantum metric for the flat-band states.

Various methods have been developed to construct exactly flat bands on a lattice~\cite{maimaiti2017compact,PhysRevB.104.085144,miyahara2005flat,kim2023general,PhysRevResearch.4.043151,cualuguaru2022general}, one of which is to use bipartite crystalline lattices (BCL), which naturally host degenerate flat bands due to their inherent chiral symmetry \cite{cualuguaru2022general}. When partitioned into sublattices $L_1$ ($N_{L_1}$ orbitals) and $L_2$ ($N_{L_2}$ orbitals) with $N_{L_1} < N_{L_2}$, the BCL Hamiltonian exhibits a block off-diagonal structure $H_{\mathrm{f}} = \sum_{\bm{k}} \Psi_{L_1,\bm{k}}^{\dagger} S_{\bm{k}} \Psi_{L_2,\bm{k}} + \mathrm{h.c.}$,
%Bipartite crystalline lattices (BCL) naturally host degenerate flat bands due to their inherent chiral symmetry \cite{cualuguaru2022general}. When partitioned into sublattices $L_1$ ($N_{L_1}$ orbitals) and $L_2$ ($N_{L_2}$ orbitals) with $N_{L_1} < N_{L_2}$, the BCL Hamiltonian exhibits a block off-diagonal structure:
%	\begin{equation}
%		H_{\mathrm{f}} = \sum_{\bm{k}} \Psi_{L_1,\bm{k}}^{\dagger} S_{\bm{k}} \Psi_{L_2,\bm{k}} + \mathrm{h.c.},
%		\label{Eq_BCL_Hamilt}
%	\end{equation}
where $S_{\bm{k}}$ is a $N_{L_1} \times N_{L_2}$ matrix, and $\Psi_{L_1,\bm{k}}$($\Psi_{L_2,\bm{k}}$) is a $N_{L_1}$($N_{L_2}$)-dimensional spinor. This structure guarantees $N_f = N_{L_2} - N_{L_1}$ degenerate flat bands at zero energy. To readily engineer flat-band quantum geometry, we further partition $L_2$ into $L_{2,A}$ ($N_{L_1}$ orbitals) and $L_{2,B}$ ($N_f$ orbitals), and confine the minimal coupling $H_{\text{c}}$ to $L_{2,B}$. Accordingly, $S_{\bm{k}}$ is partitioned as $ S_{\bm{k}} = (S_{1,\bm{k}}, S_{2,\bm{k}})$, with $S_{1,\bm{k}}$ ($S_{2,\bm{k}}$) being the size of $N_{L_{1}} \times N_{L_1}$ ($N_{L_{1}} \times N_{L_f}$). The flat-band eigenvectors $\mathcal{F}_{\bm{k},\alpha}$ ($N_{L_1} < \alpha \le N_{L_2}$) reside exclusively on $L_2$ and take the form
\begin{equation}
    \mathcal{F}_{\bm{k},\alpha} = 
    \begin{pmatrix}
        0 \\ \bm{\phi}_{1,\bm{k},\alpha} \\ \bm{\phi}_{2,\bm{k},\alpha}	
    \end{pmatrix},
    \label{Eq_BCL_FB_eigenvectors2}
\end{equation}
with components satisfying the eigenvalue equation $S_{1,\bm{k}} \bm{\phi}_{1,\bm{k},\alpha} + S_{2,\bm{k}} \bm{\phi}_{2,\bm{k},\alpha} = 0$. The quantum geometry of the flat bands can be engineered by imposing specific constraints on a quantum geometry indicator $\lambda_{\bm{k}}$ (see Methods). By imposing the {\it local atomic constraint} $\lambda_{\bm{k}} \ll 1$, we find that each flat-band eigenvector can be locally (e.g., the $\Gamma$-point) identified with a single atomic orbital, with $ \mathcal{F}_{\bm{k} \rightarrow 0, \alpha} \approx 	(0, \overset{\overset{(2N_{L_1}+\alpha) \text{th}}{\uparrow}}{...\ , 1,...,}\ 0)^{\operatorname*{T}}$. Therefore, near the $\Gamma$-point, we introduce the minimal topological coupling among the flat bands [suppose $N_f = 2^g$ ($g \in \mathbb{Z}^+$)], and the total Hamiltonian becomes
\begin{equation}
    \begin{aligned}
        &H_{\mathrm{t}}(\bm{k}) = \begin{pmatrix}
            0	 	& S_{1,\bm{k}} & S_{2, \bm{k}} \\
            S_{1,\bm{k}}^{\dagger}  & 0  & 0\\
            S_{2, \bm{k}}^{\dagger}  &0   & H_{d}^{(g)}(\bm{k})
        \end{pmatrix} \\
        \text{with}  
        & \quad H_d^{(g)}(\bm{k}) = \sum_{i=1}^d v_i k_i \gamma_i^{(g)} + m \gamma_{d+1}^{(g)},
        \label{Eq_topo_FB_Hamilt}
    \end{aligned}
\end{equation}
where $d$ is the spatial dimension, and $2^g \times 2^g$ Gamma matrices $\{\gamma_i^{(g)}\}$ generate the Clifford algebra $\mathrm{Cl}(2g+1,0)$. The Dirac form $H_d^{(g)}(\bm{k})$ provides maximal flexibility for realizing different Altland-Zirnbauer symmetry classes \cite{RevModPhys.88.035005} with dimension $d < 2g+1$. Projecting $H_{\mathrm{t}}(\bm{k})$ onto the flat-band subspace spanned by $\mathcal{F}_{\bm{k}} = (\mathcal{F}_{\bm{k},N_{L_1}+1},\mathcal{F}_{\bm{k},N_{L_1}+2},...,\mathcal{F}_{\bm{k},N_{L_2}})$, we can obtain the effective model
	\begin{equation}
		H^{\mathrm{eff}}(\bm{k}) = \mathcal{F}_{\bm{k}}^{\dagger} H_{\mathrm{t}} \mathcal{F}_{\bm{k}} \xrightarrow{\bm{k}\to 0} H_d^{(g)}(\bm{k}),
		\label{Eq_effective_model}
	\end{equation}
demonstrating that the low-energy physics is directly inherited from $H_d^{(g)}(\bm{k}\to 0)$. For instance, the mass term $m$ controls the topological transitions, while the velocities $v_i$ set the bandwidth scale.

Furthermore, by imposing additional constraints for the quantum geometry indicator at a certain point where the gap separating flat bands and dispersive bands is located, e.g., the $X$-point, we find that the $xx$-component of the non-Abelian quantum metric for the flat bands at $X$-point can significantly exceed the lattice scale (see Methods), $\mathcal{G}^{\alpha\alpha}_{xx}({\bm{k}}_X)  \gg a^2$,
%\begin{equation}
%    \mathcal{G}_{xx}({\bm{k}}_X)  \gg a^2.
%\end{equation}
and it can be further enhanced by reducing the band gap. In contrast, the $\Gamma$ point does not exhibit large non-Abelian quantum metric due to the local atomic constraint.
Therefore, a quantum metric hot spot near $X$ in the Brillouin Zone (BZ) is expected to emerge. As a remark, the BZ distribution of the non-Abelian quantum metric for the flat bands later determines the QML at different momentum channels, profoundly shaping the spatial behaviors of the TBMs.

%%%%%%%%%%%%%%%%%%%%%%%%%%%%%%%%%

\subsection{\label{Sec_QML_TBM} Manifested QML in the TBMs}

Multi-band TBMs in general can be expressed as linear combinations of hybrid Wannier orbitals $\mathcal{W}_{\alpha}(x-R_x,\tilde{\bm{k}})$ located at different lattice sites, modulated by a vector $\mathcal{U}_{\alpha}(R_x, \tilde{\bm{k}})$:
\begin{equation}
    \Psi^{\text{B}}(x, \tilde{\bm{k}}) 
    = \sum_{\alpha} \sum_{R_x} \mathcal{U}_{\alpha}(R_x, \tilde{\bm{k}}) \mathcal{W}_{\alpha}(x-R_x,\tilde{\bm{k}})
    \label{Eq_edge_states_wave_functions},
\end{equation}
where we have imposed open boundary conditions (OBC) along the \( x\)-direction so that $\tilde{\bm{k}} = (k_y,...,k_d)$ remains a good quantum number. The envelop vector $\mathcal{U}_{\alpha}(R_x, \tilde{\bm{k}})$ is an eigenvector of the flat-band effective Hamiltonian in Eq.~(\ref{Eq_effective_model}) with the transformation $k_x \rightarrow -i\partial_{R_x}$.
%, and is independent of $\mathcal{W}_{\alpha}(x-R_x,\tilde{\bm{k}})$ only when the Wannier basis is optimally localized.
%in general depends on the choice of $\mathcal{W}_{\alpha}(x-R_x,\tilde{\bm{k}})$. 
It typically takes an oscillatory decay with length scale $\xi_c$, which is entirely driven by bare band dispersion when the Wannier basis is optimally localized (see Supplementary Note II A). On the other hand, the Wannier basis in Eq.~(\ref{Eq_edge_states_wave_functions})---Fourier transformed from the flat-band eigenvectors in Eq.~(\ref{Eq_BCL_FB_eigenvectors2})---contains the information of the quantum geometry. In the limit $\xi_c \rightarrow 0$, the Wannier orbitals at the boundary site form observable eigenstates for the system, which naturally adopts an exponential decay due to the translation symmetry~\cite{PhysRev.115.809,PhysRevLett.86.5341}. Therefore, taking a specific form for the exponentially localized Wannier orbitals with length scale $\xi_g$, we find that the TBMs have a simple form
\begin{equation}
    \begin{aligned}
        \Psi^{\text{B}}(x,\tilde{\bm{k}}) 
        = A_c e^{-x/{\xi_c}} + A_{g} e^{-x/{\xi_g}},
    \end{aligned}
    \label{Eq_edge_states_wave_functions2}
\end{equation}
where the coefficients $A_c$ and $A_{g}$, implicitly dependent on $\tilde{\bm{k}}$, are determined by the details of the Wannier orbitals (see Supplementary Note II C).
%and the conventional length .
%The form of the boundary modes in Eq.~(\ref{Eq_edge_states_wave_functions2}) exactly reproduces the Majorana zero mode wave functions derived for the Lieb-Kitaev model \cite{guo2024majoranazeromodesliebkitaev}. 
Notably, the general form of Eq.~(\ref{Eq_edge_states_wave_functions2}) is not tied to any specific model. This suggests that multi-band TBMs universally exhibit two phases of behaviors, namely, a conventional behavior and a geometric behavior, corresponding to the two terms in Eq.~(\ref{Eq_edge_states_wave_functions2}). Near the boundary, the states initially undergo a conventional oscillatory decay with a diminishing oscillation amplitude. Subsequently, %once the oscillation fades, 
a pure exponential decay emerges. The competition between them depends on the relative scale of $\xi_c$ and $\xi_g$.

In particular, in the flat-band limit when $\xi_c \rightarrow 0$, we find that the spread for the TBMs, $\Omega^{x}_{\Psi_{\alpha}^{\text{B}}} \equiv \int_{x} (\left(\Delta{\hat{x}} \right)_{|\Psi_{\alpha}^{\text{B}}(x)|^2}^2 \diff x$, satisfies the following inequality (see Supplementary Note II D):
\begin{equation}
    \Omega^{x}_{\Psi_{\alpha}^{\text{B}}} \ge  \frac{a}{2\pi} \sum^{N_f}_{\alpha \alpha'} \int_{\text{BZ}}  {u}_{\alpha}^{*} \mathcal{G}^{\alpha\alpha'}_{xx}(\bm{k}) {u}_{\alpha'} \diff k_x,
    \label{Eq_inequality_LocalizationFunc_ge_nonAbelian_QM}
\end{equation}
where $\mathcal{G}^{\alpha\alpha'}_{xx}(\bm{k})$ is the non-Abelian quantum metric of the original $N_f$ degenerate flat bands, and $|u\rangle=(u_1,...,u_{N_f})$ is an eigenvector of $\gamma^{g}_i \gamma^{g}_{d+1}$ (e.g., in the case of 2 topological flat bands with $x$-OBC in 2D, one has $g=1$, $i=1$, $d=2$, the eigenvector of $\gamma^{g}_i \gamma^{g}_{d+1} = \sigma_1\sigma_3$ is $|u\rangle=(1,\pm i)^T/\sqrt{2}$). The BZ-integrated non-Abelian quantum metric on the right-hand side of Eq.~(\ref{Eq_inequality_LocalizationFunc_ge_nonAbelian_QM}) sets a lower bound for the TBMs spread, which naturally defines a length scale, namely, the QML~\cite{hu2025anomalous,guo2024majoranazeromodesliebkitaev,li2024flatbandjosephsonjunctions}. A general QML is defined with respect to some localizing direction (e.g., the OBC-direction) $\hat{i}$ with a certain momentum channel $\tilde{\bm{k}}$, as given in Eq.~(\ref{Eq_QML_definition}). This also aligns with the QML definition in Ref.~\cite{guo2024majoranazeromodesliebkitaev}, as discussed in Supplementary Note II D.

%%%%%%%%%%%%%%%%%%%%%%%%%%%%%%%%%%%%%%%%
\begin{figure}[htbp]
    \centering
    \includegraphics[width=1\linewidth]{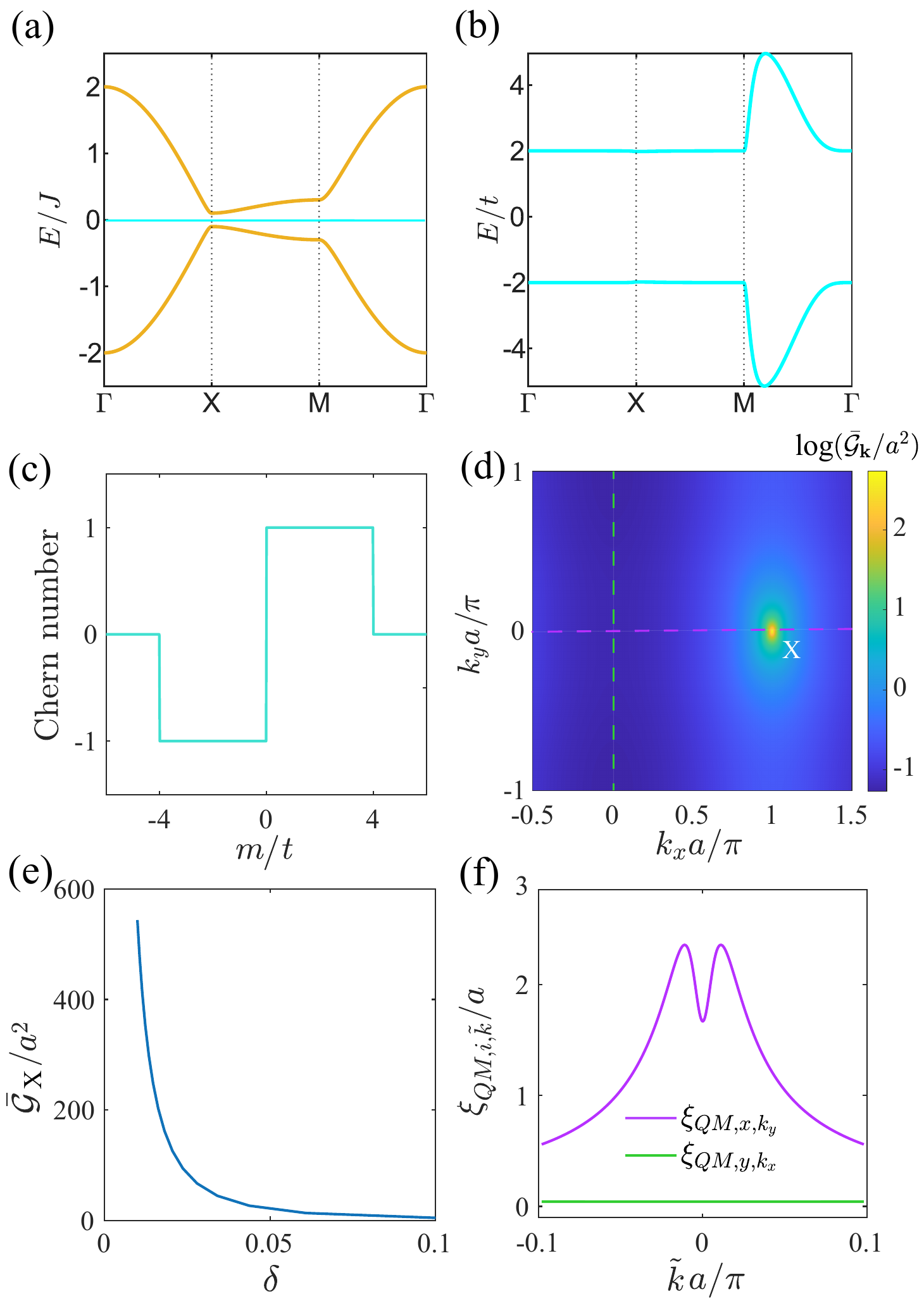}
    \caption{ \textbf{Band structure, Topology and QML of the Lieb-QWZ model.} (a) Band structure of the Lieb-QWZ model along high-symmetry points.
    (b) Zoomed-in view of the two topological flat bands. 
    (c) The Chern number evolution of a flat band as a function of the mass term $m/t$. 
    (d) The BZ distribution of the traced non-Abelian quantum metric tensor $\bar{\mathcal{G}}_{\bm{k}} \equiv \frac{1}{N_f}\sum^{N_f}_{\alpha\alpha'} \operatorname{Tr}[{\mathcal{G}_{ab}^{\alpha\alpha'}}(\bm{k})]$ of the exactly degenerate flat bands, featuring a hot spot at the $X$-point. The green (purple) dashed line represents $k_x=0$ ($k_y=0$) cut lines respectively.
    (e) The $\bar{\mathcal{G}}_{X}$ as a function of $\delta$ $(\propto \Delta_X)$. 
    (f) The \(x\)(\(y\))-directional QML $\xi_{QM,x,k_y}$ ($\xi_{QM,y,k_x}$) around the $\Gamma$-channel. %The parameters for (d-f) are the same as (a), except $m=\tilde{t}=t=0$, $\alpha = 0.5$, $\delta = 0.01$. 
    Parameters: $(m,\tilde{t},t,\alpha,\delta)=(4\times 10^{-4},2\times 10^{-4},2\times 10^{-4},0.1,0.05)$ for (a) to (c); $(m,\tilde{t},t,\alpha,\delta)=(0,0,0,0.5,0.01)$ for (d) to (f). $J=1$ for all panels.
    }
    \label{Fig_Lieb_QWZ_lattice_band_QM}
\end{figure}

\subsection{\label{sec_Models} Topological flat band models}
%The simplest topological flat band model to exemplify our framework can be implemented on a variant-Lieb lattice.
We exemplify our framework by constructing topological flat bands on a variant Lieb lattice. Specifically, we extend sublattice C of the original Lieb lattice~\cite{PhysRevLett.62.1201} into a set of $2^g$ ($g\in \mathbb{Z}^+$) equivalent sublattice replicas, giving rise to $2^g$ degenerate flat bands. By incorporating a coupling matrix $H_d^{(g)}(\bm{k})$ among the extended sublattices, we can realize various topological classes depending on the choice of $d$ and $g$. The Hamiltonian, as defined in Eq.~\eqref{Eq_topo_FB_Hamilt}, is given by
\begin{equation}
    \begin{aligned}
        & S_{1,\bm{k}} = J[(1+\delta)+(1-\delta)e^{ik_1a} ]\\
        & S_{2, \bm{k}} = \alpha J \left[(1+\delta) - (1-\delta)e^{ik_2a}\right] \mathcal{I}_{1\times 2^g}
        \label{Eq_S1_S2_Lieb_TI}.
    \end{aligned}
\end{equation}
Here, the parameter $\delta$ ($0<\delta \ll 1$) controls the staggered hoppings between unit cells, determining the band gap and the non-Abelian quantum metric of the flat bands. The parameter $\alpha$ ($0< \alpha \lesssim 1$) shapes the energy spectrum of the flat bands.
%controls $\lambda_{\bm{k}}$, defined in Eq.~(\ref{Eq_quantum_geometry_indicator}), 
%shapes the flat band structure to approximate that of $H_d^{(g)}(\bm{k})$. Smaller $\alpha$ expands the BZ region where the flat bands overlap with the bare $H_d^{(g)}(\bm{k})$ bands, while at $\alpha = 1$, overlap occurs only at the $\Gamma$-point.
As expected from our framework, the variant Lieb lattice exhibits nontrivial quantum metric. Specifically, in terms of the local quantum geometry indicator (see Methods), at the $\Gamma$-point, $\lambda_{\Gamma} = \alpha\delta \ll 1$, which satisfies the local atomic constraint. In contrast, at the $X$-point where the direct band gap is located, one has $\lambda_{X} \sim 1$. This results in a peak for the quantum metric distribution in the BZ, controllable by tuning the band gap via $\delta$, as demonstrated in the subsequent models.

%%%%%%%%%%%%%%%%%%%%%%%%%  
\begin{figure*}[htbp]
    \centering
    \includegraphics[width=0.9\linewidth]{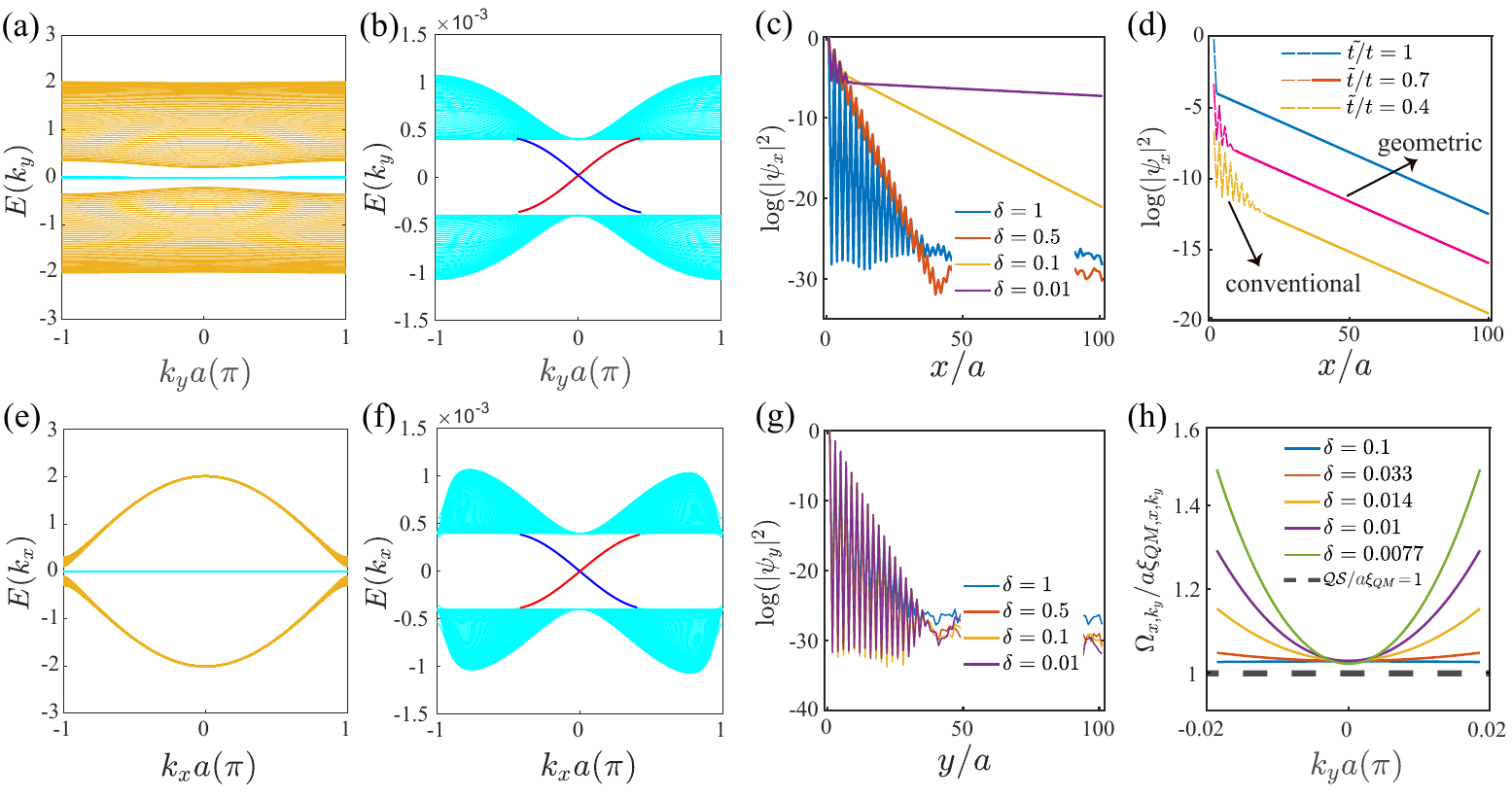}
    \caption{ \textbf{Spatial behaviors of TBMs in the Lieb-QWZ model.} (a,b,e,f) The spectrum for the Lieb-QWZ model with OBC (a) in the \(x\)-direction and (e) in the \(y\)-direction. (b) and (f) shows the corresponding flat bands zoomed in, with in-gap modes highlighted in red and blue for the left and right branches, respectively.
    %The parameters are $J=1$, $\tilde{t}= 2\times10^{-4}J$, $m = 2\tilde{t}$, $t = \tilde{t}$, $\delta = 0.05$ and $\alpha = 0.1$ (The parameters will be the same in the following unless otherwise specified).
    (c,g) The logarithmic wave functions for different $\delta$ with OBC in (d) the \(x\)-direction and (h) the \(y\)-direction.
    To manifest both behaviors, we keep the parameters the same as before except $t = 0.5\tilde{t}$. 
    (d) The logarithmic wave functions for different $\tilde{t}/t$ with OBC in the \(x\)-direction (vertically shifted for visual clarity).
    (h) In the case $t=\tilde{t}$, the ratio $\Omega_{x,k_y}/a\xi_{QM}$ (where $a$ is the lattice constant) is consistently bounded below by $1$ in the Lieb-QWZ model, in excellent agreement with Eq.~(\ref{Eq_inequality_LocalizationFunc_ge_nonAbelian_QM}). Parameters: $t = \tilde{t}$ for (a,b,e,f), and $J=1$, $\tilde{t}= 2\times10^{-4}J$, $m = 2\tilde{t}$, $\delta = 0.05$ and $\alpha = 0.1$ for all panels.}
\label{Fig_Lieb_QWZ_OBC_spectrum_wavefunc_QSQM}
\end{figure*}

%%%%%%%%%%%%%%%%%%%%%%%%%%%%%%%%%%

%\subsubsection{\label{sec_3-2-1} Lieb-QWZ model}
As a concrete example, taking $g = 1$, $d = 2$, the $2\times 2$ block is well known as the Qi-Wu-Zhang (QWZ) model which is a prototypical lattice model for Chern insulators~\cite{PhysRevB.74.085308}:
\begin{equation}
    H_2^{(1)}(\bm{k}) =  2t\sin(k_1a)\sigma_x + 2t\sin(k_2a)\sigma_y + M(\bm{k}) \sigma_z,
    \label{Eq_QWZ_hamilt}
\end{equation}
with $M(\bm{k}) = m - 2\tilde{t}(\cos(k_1a) + \cos(k_2a))$. The flatness condition requires $\operatorname{max}(t, \tilde{t}, m) \ll \alpha \delta J$. By substituting Eqs.~\eqref{Eq_S1_S2_Lieb_TI}, \eqref{Eq_QWZ_hamilt} into Eq.~\eqref{Eq_topo_FB_Hamilt}, we embed the QWZ model within a variant-Lieb lattice, yielding the four-band Lieb-QWZ model. Figure~\ref{Fig_Lieb_QWZ_lattice_band_QM}(a) displays the full band structure, where two dispersive bands sandwich a pair of flat bands in the middle, with a zoomed-in view in Fig.~\ref{Fig_Lieb_QWZ_lattice_band_QM}(b). 
The Lieb-QWZ model belongs to the class A in the AZ classification~\cite{RevModPhys.88.035005}, characterized by the Chern number. In Fig.~\ref{Fig_Lieb_QWZ_lattice_band_QM}(c), we compute the Chern number of one flat band as a function of $m/t$, revealing a topological nontrivial regime for $|m|<4t$.
The flat bands inherit the quantum metric from the original degenerate manifolds, which can be recovered by turning off $H_d^{(g)}$ (i.e., $t=\tilde{t}=m=0$).
We then compute the traced non-Abelian quantum metric for the degenerate flat bands, defined as $\bar{\mathcal{G}}_{\bm{k}} \equiv \frac{1}{N_f} \operatorname{Tr}\left[ \sum^{N_f}_{\alpha\alpha'} {\mathcal{G}_{ab}^{\alpha\alpha'}}(\bm{k})\right]$, and plot its distribution across the BZ in Fig.~\ref{Fig_Lieb_QWZ_lattice_band_QM}(d). The quantum metric peaks prominently near the $X$-point, as expected.
Figure~\ref{Fig_Lieb_QWZ_lattice_band_QM}(e) shows the traced non-Abelian quantum metric $\bar{\mathcal{G}}_{X}$ at the $X$-point, demonstrating enhancement by reducing the band gap $\Delta_{X}$. 

Integrating the non-Abelian quantum metric over the cut line along different $k_y$ ($k_x$)-channels according to Eq.~\eqref{Eq_QML_definition}, we obtain the orientation-dependent QML $\xi_{QM,x,k_y}$ ($\xi_{QM,y,k_x}$) around the $\Gamma$-channel in Fig.~\ref{Fig_Lieb_QWZ_lattice_band_QM}(f), exhibiting an anisotropic feature. Specifically, the \( x \)-directional QML $\xi_{QM,x,k_y=0}$, integrated over $k_x$ along $k_y=0$, intersects the $X$-point, acquiring a finite and tunable value, as indicated by the purple dashed line in Fig.~\ref{Fig_Lieb_QWZ_lattice_band_QM}(d). In contrast, the \(y\)-directional QML, $\xi_{QM,y,k_x=0}$---integrated over $k_y$ along $k_x=0$ while bypassing $\bm{k}_X$---remains zero and independent of $\delta$, as shown by the green dashed line in the same figure. 
%Consequently, the \(x\)-directional TBMs exhibit geometric behavior tunable via $\delta$, while the \(y\)-directional TBMs do not.
The Lieb-QWZ model provides a minimal topological model with tunable non-Abelian quantum metric. Additional topological flat band models, belonging to distinct topological classes, are constructed in Supplementary Note III to further illustrate our framework.

\subsection{Boundary-modes localization in the Lieb-QWZ model}
In the topological regime of the Lieb-QWZ model, in-gap TBMs emerge within the flat bands under $x$ ($y$)-OBC, as shown in Fig.~\ref{Fig_Lieb_QWZ_OBC_spectrum_wavefunc_QSQM}(b) [(f)], which are zoomed in from the full spectrum in Fig.~\ref{Fig_Lieb_QWZ_OBC_spectrum_wavefunc_QSQM}(a) [(e)].
%Topological boundary modes emerge under OBC.
%Under $x$ and $y$-OBC, the spectrum of the Lieb-QWZ is shown in Fig.~\ref{Fig_Lieb_QWZ_OBC_spectrum_wavefunc_QSQM}(a) and (e) respectively, with zoomed-in flat bands in Fig.~\ref{Fig_Lieb_QWZ_OBC_spectrum_wavefunc_QSQM}(b) and (f).
As clarified above, multi-band TBMs typically exhibit two sequential phases of spatial behaviors: A conventional oscillatory decay driven by bare band dispersion, and a geometrical exponential decay from the localized Wannier orbitals controlled by the QML. This is exactly what we observe in Fig.~\ref{Fig_Lieb_QWZ_OBC_spectrum_wavefunc_QSQM}(d),
where the conventional and geometrical behaviors are denoted in dashed and solid lines, respectively.
However, depending on the relative scales of $\xi_c$ and $\xi_g$ in Eq.~\eqref{Eq_edge_states_wave_functions2}, these two behaviors may not always be observed simultaneously. When the conventional length dominates, i.e., $\xi_c \gg \xi_g > 0$, it is found that the first term in Eq.~\eqref{Eq_edge_states_wave_functions2} dominates the second one, effectively overshadowing the geometrical behavior. This is well reflected in Fig.~\ref{Fig_Lieb_QWZ_OBC_spectrum_wavefunc_QSQM}(c). By increasing $\delta$, the geometrical length $\xi_g$ for the \(x\)-directional TBMs diminishes, and eventually becomes invisible. In contrast, the TBMs under the \( y \)-OBC are insensitive to $\delta$ and only exhibit the conventional behavior, as shown in Fig.~\ref{Fig_Lieb_QWZ_OBC_spectrum_wavefunc_QSQM}(g). This indicates that the \(y\)-directional geometrical length is much smaller than the conventional length: $\xi_g \ll \xi_c$. Their distinct behaviors are consistent with the anisotropic QML we found in Fig.~\ref{Fig_Lieb_QWZ_lattice_band_QM}(f).
%%%%%%% Add comparison of calculated QML in both directionstilde{t}/t$ is, the smaller $\xi_c$ is, the sooner geometric behavior occurs. When $\tilde{t}/t=1$ with $\xi_c = 0$, it only showcase the geometric behavior.

Conversely, the geometrical length will manifest itself when $\xi_g > \xi_c$ and become the dominant length scale when $\xi_g \gg \xi_c > 0$. Examining the two terms in Eq.~(\ref{Eq_edge_states_wave_functions2}) in this regime, it is evident that in the long range, the boundary modes are ultimately governed by $\xi_g$. However, over the first few sites near the boundary, the conventional decay may dominate, and soon transitions to a slower exponential decay dictated by $\xi_g$. The conventional length $\xi_c$ determines how quickly this transition occurs: a large $\xi_c$ delays the transition, as shown in Fig.~\ref{Fig_Lieb_QWZ_OBC_spectrum_wavefunc_QSQM}(d). When $\xi_c \rightarrow 0$, the TBMs only follow the geometrical behavior. 

To verify the universal bound from the QML in Eq.~(\ref{Eq_inequality_LocalizationFunc_ge_nonAbelian_QM}), we compute the \(x\)-OBC spread function $\Omega_{x, k_y}$ of the in-gap modes near the $\Gamma$-channel ($\tilde{\bm{k}}=0$), where the conventional length is negligible. The ratio $\Omega_{x,k_y}/a\xi_{QM,x,k_y}$, presented in Fig.~\ref{Fig_Lieb_QWZ_OBC_spectrum_wavefunc_QSQM}(h), clearly demonstrates that the QML sets a lower bound on the spread of flat-band TBMs, with tighter saturation for larger $\delta$.
%Away from the $\Gamma$-channel, the Lieb-QWZ model shows an increasing gap between these quantities. 
The underlying reason can be traced back to Eq.~\eqref{Eq_inequality_LocalizationFunc_ge_nonAbelian_QM}, where the dropped terms during inequality scaling diminish in the trivial quantum geometry limit (small quantum metric). Meanwhile, it is evident from Fig.~\ref{Fig_Lieb_QWZ_lattice_band_QM}(e) that the QML yields a smaller value for larger $\delta$, leading to a tighter bound. 
In particular, in the trivial atomic limit (vanishing QML), the left and right sides of Eq.~\eqref{Eq_inequality_LocalizationFunc_ge_nonAbelian_QM} both approach zero, concealing the geometric behavior.

%%%%%%%%%%%%%%%%%%%%%%%%%%%%%%%%%%%%%%%%%%%

\begin{figure}[h]
    \centering
    \includegraphics[width=1\linewidth]{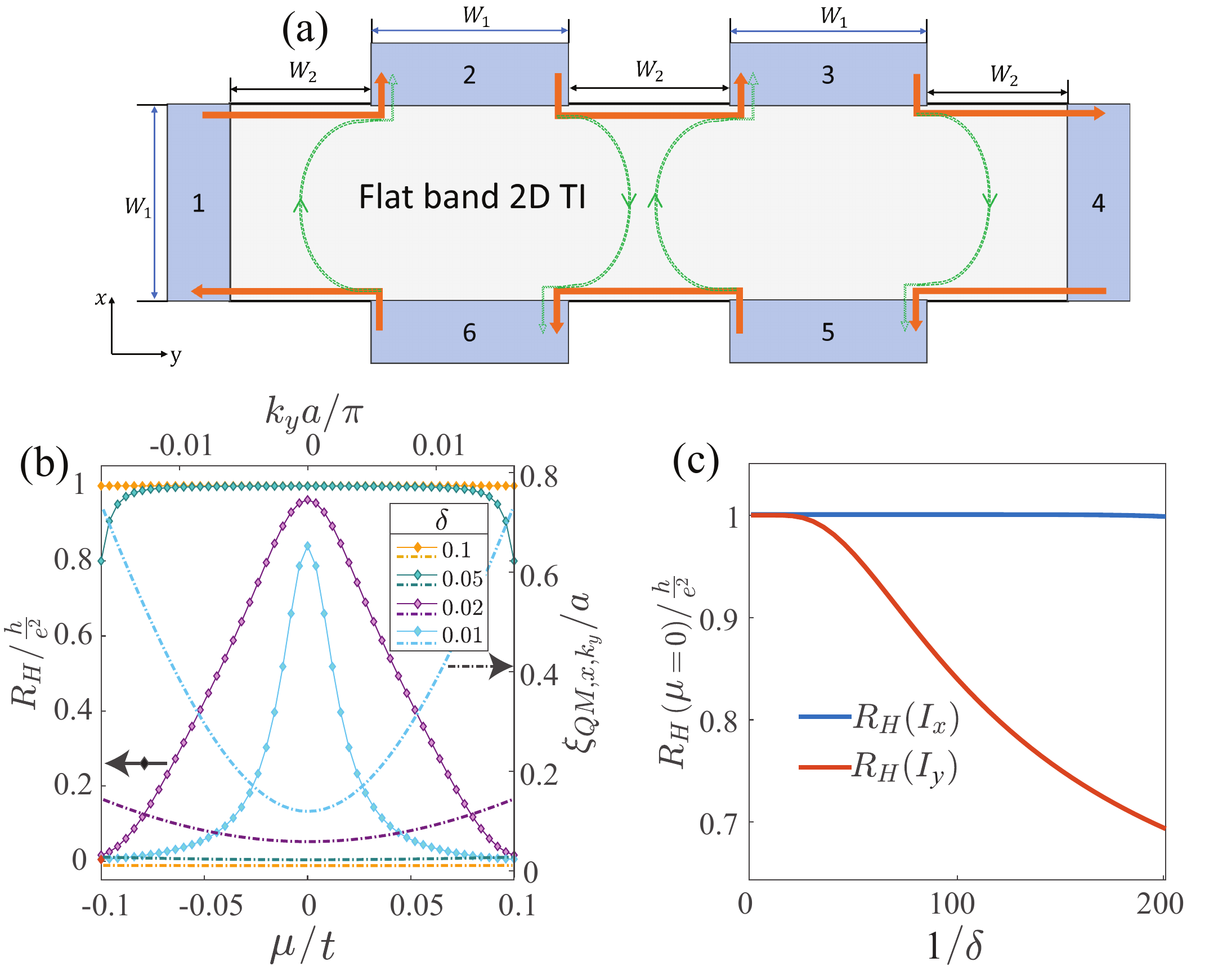}
    \caption{\textbf{Quantum Hall resistivity and the QML.} (a) A 6-terminal Hall bar device. The green dashed lines represent inter-edge scattering of the chiral edge states due to a large QML. (b) The Hall resistivity $R_H$ (solid lines) and the QML $\xi_{QM,x,k_y}$ (dash-dot lines) versus the chemical potential $\mu/t$ for different $\delta$. The chemical potential in the lower axis is locked with the momentum in the upper axis due to the linear dispersion of TBMs. The width for the device and the leads is fixed at $W_1=30a$, and the spatial separation of the terminals is set as $W_2=400a$. Other parameters are the same as Fig.~\ref{Fig_Lieb_QWZ_OBC_spectrum_wavefunc_QSQM}. (c) The zero-bias Hall resistivity versus $1/\delta$ for current driving along the \(x\)-direction and \(y\)-direction respectively. 
    The \(y\)-directional zero-bias Hall resistivity decreases with an increasing QML while the \(x\)-direction response remains quantized. }
    \label{Fig_Flatband_TI_Hall}
\end{figure}
%%%%%%%%%%%%%%%%%%%%%%%%%%%%%%%%%%%%%%%%%%%
\subsection{Quantum Hall plateau shaped by QML}
The quantum anomalous Hall effect is experimentally characterized by a quantized Hall plateau and vanishing longitudinal conductivity. The quantized plateau is robust against disorder but will deviate from perfect quantization due to finite-size effects~\cite{PhysRevB.92.081107,PhysRevLett.101.246807,chen2023chiral}. As we point out in this work, the localization behavior of the chiral edge states can be controlled via the QML. A large QML promotes coupling between the edge states on opposite edges [denoted by the green lines in Fig.~\ref{Fig_Flatband_TI_Hall}(a)]. This coupling leads to a deviation of the Hall resistivity from its quantized value $h/e^2$.

Since gapless chiral edge states with linear dispersion exhibit energy-momentum locking, the in-gap Hall resistivity around zero energy serves as a marker for the QML. We now examine how the Hall plateau is affected by the QML using a 6-terminal Hall-bar device. As illustrated in Fig.~\ref{Fig_Flatband_TI_Hall}(a), a current $I_y$ is driven along the $y$-direction from terminal 1 to terminal 4. Four additional terminals, each with a width equal to that of the device ($W_1$), are arranged on opposite edges and separated by a length $W_2$ from each other. By measuring the voltage difference between terminal 2 and 3, $V_{23}=V_2-V_3$, and the voltage difference between terminal 2 and 6, $V_{26}=V_2-V_6$, we obtain the longitudinal and Hall resistivity as
	\begin{equation}
		\begin{split}
			R_{xx} = \frac{V_{23}}{I_x}, \quad
			R_{H} = \frac{V_{26}}{I_x}.
		\end{split}
	\end{equation}	
Taking the Lieb-QWZ model as an example, we perform the multi-terminal recursive Green's function method~\cite{datta1997electronic} to simulate the Hall response. As shown in Fig.~\ref{Fig_Flatband_TI_Hall}(b), when the QML is small (large $\delta$), the chiral edge states at opposite edges remain well-separated, yielding a perfectly quantized Hall plateau (orange solid line). When we increase the QML by reducing $\delta$, the Hall resistivity drops below the quantized value due to a stronger overlap between TBMs on opposite edges.
%%%%%%%%%%%
And the resistivity also varies with chemical potential $\mu$. This deviation is shaped by the BZ-distribution of the QML, particularly near the $\Gamma$-channel where the conventional behavior vanishes. Specifically, since the in-gap TBMs exhibit linear dispersions with Fermi velocity $v_{\Gamma} = 2ta/\hbar$, the momentum range, e.g., $k_ya \in (-0.015\pi, 0.015\pi)$, is locked to an energy range of $\mu \in (-0.1t, 0.1t)$. Consequently, the QML distribution in the momentum space is projected onto the energy space, which is reflected in the Hall resistivity measurements. The shape of the quantum Hall plateau around zero-energy (solid lines) in Fig.~\ref{Fig_Flatband_TI_Hall}(b) mirrors the inverse QML distribution around $\Gamma$-channel (dash-dot lines), in which a dip at $k_y=0$ indicates a smaller QML and thus a poorer coupling between chiral edge states, leading to a larger Hall resistivity.
%%%%%%%%%%%
Focusing on the zero-energy response (the $\Gamma$ channel in momentum space), we plot the Hall resistivity as a function of $1/\delta$ in Fig.~\ref{Fig_Flatband_TI_Hall}(c). As indicated by the red line, $R_H(I_y)$ decreases with increased $\xi_{QM,x, k_y=0}$ (via $1/\delta$), reflecting stronger couplings between the chiral edge states.

To further verify how the directional properties of the QML affect the Hall responses, we drive a current along the $x$-direction and evaluate the Hall resistivity $R_H(I_x)$ upon varying the QML, finding that the quantization of $R_H(I_x)$ is robust, as shown by the blue line in Fig.~\ref{Fig_Flatband_TI_Hall}(c). This is consistent with the anisotropic QML of the Lieb-QWZ model presented in Fig.~\ref{Fig_Lieb_QWZ_lattice_band_QM}(f), in which $\xi_{QM,y, k_x=0}$ is negligibly small and nearly independent of $\delta$. Thus, when the current is driven along the \(x\)-direction, the \(y\)-direction profiles of the chiral edge states remain localized and separated, sustaining quantized Hall resistivity. 
%This contrasts sharply with $R_H(I_y)$ with nontrivial $\xi_{QM, x, k_y=0}$, highlighting the anisotropic QML behavior of the Lieb-QWZ model. 
The suppressed Hall plateau clearly reflects the QML of the in-gap TBMs, allowing indirect detection for the flat band quantum geometry.  

\subsection{QML-enabled $2\Phi_0$-$\Phi_0$ crossover in Fraunhofer pattern}

\begin{figure}[h]
    \centering
    \includegraphics[width=0.9\linewidth]{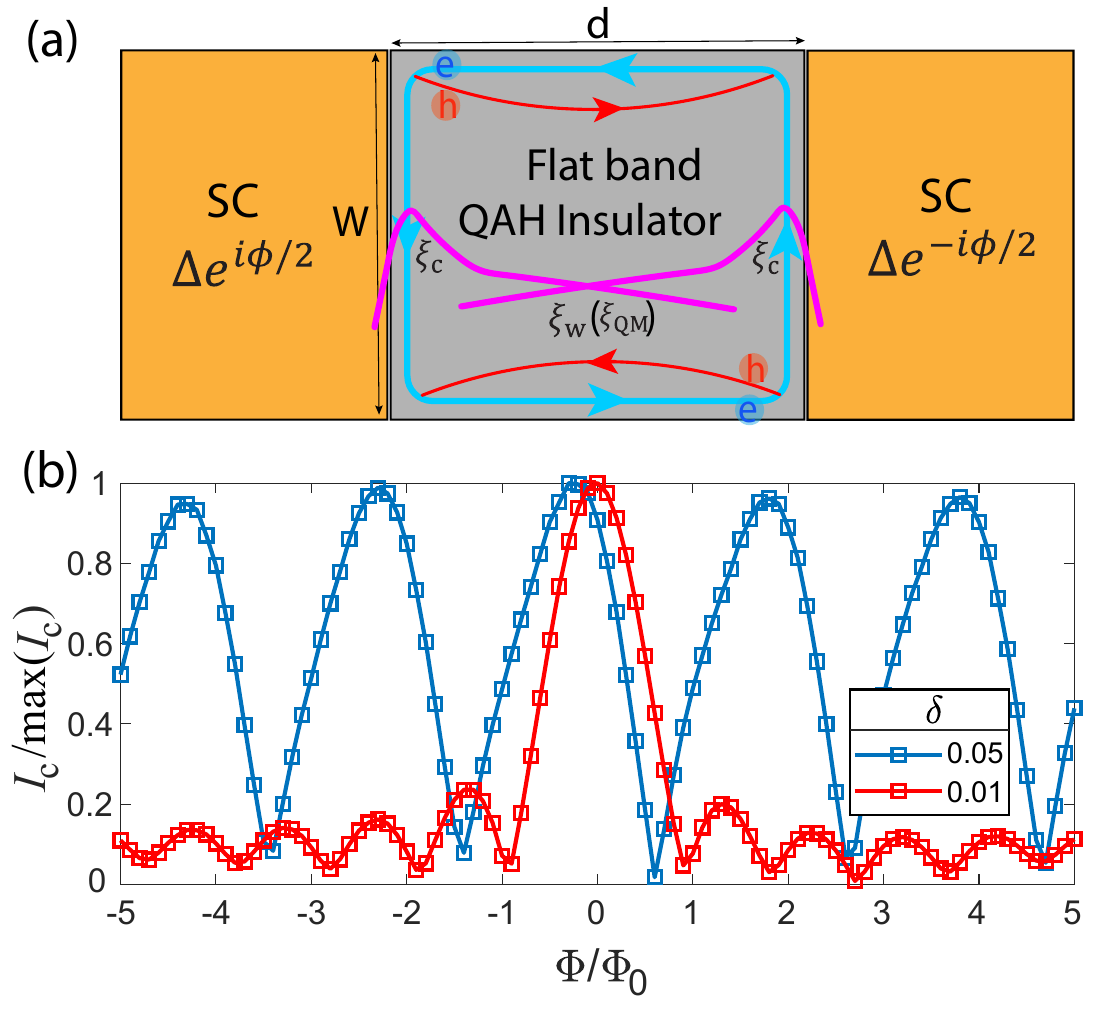}
    \caption{\textbf{Fraunhofer pattern of flat-band QAH Josephson junctions.} (a) An illustration of a flat-band QAH Josephson junction with chiral Andreev edge states coupled to each other controlled by the QML, allowing local Andreev reflections in the junction. (b) A $2\Phi_0$-$\Phi_0$ crossover in the Fraunhofer pattern occurs when increasing the QML by reducing $\delta$. The weak link has length $d=100a$ and width $W = 30a$ and is composed of the Lieb-QWZ lattice with $\alpha=0.5$, $\tilde{t} = 0.8t$. %The superconductors have bandwidth $t_s = t$, pairing gap $\Delta_s = 0.15t_s$ temperature $T=0.1\Delta_s$, and the coupling strength between the superconductors and weak link is $t_c = 0.7t$. Other parameters are the same as Fig.~\ref{Fig_Lieb_QWZ_OBC_spectrum_wavefunc_QSQM}.
    }
    \label{Fig_FP_crossover}
\end{figure}

The critical supercurrent in a Josephson junction oscillates with applied out-of-plane magnetic flux, widely known as the Fraunhofer pattern~\cite{PhysRevB.83.195441}. Of particular interest is that when the weak link is a QAH insulator, the chiral Andreev edge states mediate the supercurrent. In this case, it has also been pointed out that a crossover from $2\Phi_0$ ($\Phi_0 = h/2e$) to $\Phi_0$ in the periodicity of the Fraunhofer pattern is caused by edge-state coupling~\cite{PhysRevLett.133.056601}. Practically, this crossover occurs when the junction length $d$ is comparable to the conventional length scale $\xi_c$. In the two opposite limits, the Fraunhofer pattern exhibits $2\Phi_0$ (long junction with $d \gg \xi_{c}$) and $\Phi_0$ (short junction with $d \ll \xi_{c}$) oscillations, corresponding to the supercurrent driven by the local Andreev reflection and crossed Andreev reflection processes respectively \cite{PhysRevLett.133.056601}. 

In contrast, in flat-band QAH Josephson junctions with large QML, this crossover is expected to happen even in the conventionally long junction limit ($d \gg \xi_c$). As illustrated in Fig.~\ref{Fig_FP_crossover}(a), the chiral Andreev states at the left and right edges extend deeply into the bulk of the weak link with long-range characteristic length $\xi_g$ controllable by $\xi_{\text{QM}}$. When $\xi_g$ surpasses the junction length $d$ with $\xi_g \gg d$, even though the conventional length $\xi_c$ might be significantly shorter than $d$, the local Andreev reflection process is still expected to dominate, leading to a $\Phi_0$ Fraunhofer oscillation.
%In this case, the BCS coherence length would be replaced by the geometric length $\xi_g$ controlling the spatial extent of the edge states along the weak link.
	
We numerically calculate the Josephson supercurrent using the recursive Green's function method. 
%	The supercurrent can be obtained by:
%	\begin{equation}
%		I_s(\varphi, \Phi) = -\frac{2ek_BT}{\hbar} \frac{\diff}{\diff \varphi}\sum_{n=0}^{\infty} \ln{\det\left[1-S_A(i\omega_n) S_N(i\omega_n)\right]}
%	\end{equation}
%	where $S_A$ and $S_N$ are the scattering matrices describing the Andreev reflection at the left (right) edges and normal transmission at the top (bottom) edges, respectively. And $\omega_n = (2n+1)\pi k_BT$ is the fermionic Matsubara frequency, T is the temperature. 
As shown in Fig.~\ref{Fig_FP_crossover}(b), the critical supercurrent oscillates with the magnetic flux $\Phi$ forming the Fraunhofer pattern. When the QML is short ($\delta=0.05$), the chiral states on the left and right edges are well separated so that the crossed Andreev reflection process dominates, resulting in a conventional $2\Phi_0$ oscillation. As the QML increases with smaller $\delta$, the QML-enabled long-range behavior of the edge states becomes prominent, and a $\Phi_0$ oscillation emerges. The Fraunhofer pattern suggests that the supercurrent is mediated by Josephson vortices due to the presence of strong inter-edge scattering.
%Josephson vortices %of the size $d \times \Phi_0/Bd$ 
%are created to circulate in the weak link.
%The Fraunhofer pattern suggests that the weak link is effectively in a metallic state due to the strong edge-states coupling. 
%It is also interesting to note that $I_c(\Phi)\neq I_c(-\Phi)$, indicating an inversion symmetry breaking in this regime~\cite{chen2018asymmetric}. 
The  $2\Phi_0$-$\Phi_0$ crossover unambiguously pins down the crucial role of the QML in transport measurements.

%%%%%%%%%%%%%%%%%%%%%%%%%%%%%%%%%%%%%%%%%%%%%%%	
\section{Discussions}
%In this work, we investigate the origin, universal role, and physical consequences of the QML for TBMs in flat-band systems with nontrivial topology. The fundamental relation between the TBMs and the quantum metric revealed in this work, exists ubiquitously in topological flat-band systems, demonstrating a more complete version of the bulk-edge correspondence. Specifically, in topological systems, the existence of boundary modes is determined by the topology of the bulk states, while their long-range spatial behavior is governed by the quantum metric of the bulk states. These two aspects correspond exactly to the real and imaginary parts of the quantum geometric tensor, underscoring the conceptual importance of this work.

The pivotal role of the QML revealed in this work suggests that with only low-energy description of topological bands one misses essential quantum geometry. While full-model treatments are computationally intensive, we show that the QML is determined by the BZ-distribution of the non-Abelian quantum metric tensor---particularly dominated by hot-spot regions near direct gaps between flat and dispersive bands---allowing a reasonably accurate QML estimation through the inclusion of the topological flat bands and their dispersive neighbors. Consequently, for moiré materials, continuum models capturing these low-lying bands can also reliably predict boundary-mode localization.

In addition to the topological flat band models studied here, 
%including the 1D TIs, 2D TIs and 2D TSC models belonging to the class A, AIII, and D in the AZ ten-fold classification. As our conclusions are expected to be general, 
future works can be explored on the QML effects in 3D TIs, topological semimetals, and high-order topological insulators. In this work, we have applied our theory to the lattice model for magic-angle twisted bilayer graphene (more details, see Supplementary Note III), suggesting that 2D moire materials with topological flat bands are plausible experimental systems to explore the physics of the QML. Furthermore, if we expand the scope to new platforms such as artificial systems, the simple flat-band topological models developed here could be readily implemented in well-controlled photonic and circuit lattices~\cite{chan2025essay}. The detection of the QML in these systems would be rather promising due to their precise tunability.

% Cons for QML 
%As a remark, while the QML brings fascinating new physical phenomenon, it may be detrimental in some situations, where the boundary modes are required to be isolated from each other, e.g., the realization of the Majorana zero modes in the Poorman's scheme \cite{PhysRevB.86.134528}. In such scenarios, one needs a flat band system with trivial QML to avoid coupling of the end modes.

%\emph{New Bulk-edge correspondence}
%The intertwined relation between the TBMs and quantum metric revealed in this work exists ubiquitously in topological flat-band systems, demonstrating a more complete version of the idea of bulk-edge correspondence. That is, in topological systems, the existence of boundary modes are determined by the topology of the bulk states, while their spatial behavior is governed by the quantum metric of the bulk states. These two aspects are exactly the real and imaginary part of the quantum geometric tensor, underscoring a conceptual importance of this work.

%%%%%%%%%%%%%%%%%%%%%%%%%%%%%%
%\newpage
\quad

%\newpage
\section{Methods}

\subsection{Quantum geometry engineering on flat bands}
The quantum geometry of the flat bands can be worked out through energy scale analysis. We denote $\mathcal{E}_{1,\bm{k}}$ ($\mathcal{E}_{2,\bm{k}}$) as the set of characteristic energy scales of $S_{1,\bm{k}}$ ($S_{2,\bm{k}}$) in Eq.~(\ref{Eq_topo_FB_Hamilt}), encompassing hopping amplitudes, on-site energies, Zeeman terms, etc.
Then a {\it local quantum geometry indicator} can be defined as:
\begin{equation}
    \lambda_{\bm{k}} \equiv  \frac{\operatorname{max}(\mathcal{E}_{2,\bm{k}})}{\operatorname{min}(\mathcal{E}_{1,\bm{k}})} 
    \sim \frac{\operatorname{max}(|\phi_{1,\bm{k},\alpha}|)}{ \operatorname{min}(|\phi_{2,\bm{k},\alpha}|)},
    \label{Eq_quantum_geometry_indicator}
\end{equation}
in which the second relation arises from the flat-band eigen-equation, and the last ratio represents orbital compositions at $\bm{k}$, reflecting the local quantum geometry of the flat bands. The ratio $\lambda_{\bm{k}}$ serves as a crucial tuning parameter throughout our construction.
For instance, we can achieve trivial local quantum geometry near the $\Gamma$-point ($\bm{k} = 0$) by imposing the {\it local atomic constraint}:
\begin{equation}
    \lambda_{\bm{k} \rightarrow 0} \ll 1
    %\lambda_{\bm{k}_{\Gamma}} \ll 1
    \label{Eq_Local_Atomic_Constraint},
\end{equation}
which means that $S_{2, \bm{k} \rightarrow 0}$ has much smaller energy scales compared to $S_{1, \bm{k} \rightarrow 0}$. From the bare BCL part of Eq.~(\ref{Eq_topo_FB_Hamilt}), it is evident that Eq.~(\ref{Eq_Local_Atomic_Constraint}) restricts the flat band eigenvectors to resemble atomic orbitals around the $\Gamma$ point: $ \mathcal{F}_{\bm{k} \rightarrow 0, \alpha} \approx 	(0, \overset{\overset{(2N_{L_1}+\alpha) \text{th}}{\uparrow}}{...\ , 1,...,}\ 0)^{\operatorname*{T}} $. A minimal topological coupling among the flat bands can be introduced with [suppose $N_f = 2^g$ ($g \in \mathbb{Z}^+$)]
\begin{equation}
    \begin{aligned}
%			& H_{\mathrm{c}}(\bm{k}) = \operatorname{diag}\left(\mathbf{0}_{N_{L_1}}, H_d^{(g)}(\bm{k})\right) \\
    & H_{\text{c}}(\bm{k}) = 
   \begin{pmatrix}
      O_{N_{L_1}\times N_{L_1}}  & O\\
      O  & H_{d}^{(g)}(\bm{k})
   \end{pmatrix} \\
        \text{with}  
        & \quad H_d^{(g)}(\bm{k}) = \sum_{i=1}^d v_i k_i \gamma_i^{(g)} + m \gamma_{d+1}^{(g)}.
    \end{aligned}
    \label{Eq_top_coupling}
\end{equation}
With Eqs.~(\ref{Eq_H_tf}), %(\ref{Eq_BCL_Hamilt}) 
and (\ref{Eq_top_coupling}), we obtain the total Hamiltonian $H_{\mathrm{t}}(\bm{k})$ in Eq.~(\ref{Eq_topo_FB_Hamilt}), and the projected effective model is given in Eq.~(\ref{Eq_effective_model}). 
%Notably, the flat band basis $\mathcal{F}_{\bm{k},\alpha}$ is now incorporated with multi-band information, preserving faithful local quantum geometry through band projection in Eq.~(\ref{Eq_effective_model}). 
Besides the minimal coupling given in Eq.~(\ref{Eq_top_coupling}), alternative coupling schemes and their comparisons can be found in Supplementary Note IV.

%A rigorous symmetry alignment between $H_{\mathrm{t}}(\bm{k})$ and $H_d^{(g)}(\bm{k})$ requires additional constraints on $S_{1,\bm{k}}$ and $S_{2,\bm{k}}$. 
%%%%%%%%%%%%%%%%%%%%%%%%%%
	
Eq.~(\ref{Eq_Local_Atomic_Constraint}) trivializes the flat band states so as to be locally akin to atomic orbitals. To maintain global nontrivial quantum geometry across the BZ, the flat band eigenvectors must deviate atomic orbitals at somewhere away from the $\Gamma$ point ($\bm{k} \centernot \to 0$), e.g., the point where the direct gap separating flat bands and dispersive bands is located, supposed to be at the $X$-point:
\begin{equation}
    \lambda_{\bm{k}_X} \sim 1
    \label{Eq_nontrivialize_FB_condition}.
\end{equation}
Given the flat-band eigenvectors in Eq.~(\ref{Eq_BCL_FB_eigenvectors2}), the conditions in Eqs.~(\ref{Eq_Local_Atomic_Constraint}) and (\ref{Eq_nontrivialize_FB_condition}) ensure nontrivial quantum distance between Bloch vectors near the $\Gamma$- and $X$-point. Additionally, the $X$-point in general possesses a large non-Abelian quantum metric.
%as the quantum metric in general peaks at flat-band gap points due to the small denominator.
This can be seen by analyzing the inter-band velocity. Consider a dispersive band eigenvector $\mathcal{D}_{\bm{k}, \alpha'}=(\varphi_{1,\bm{k}, \alpha'}, \varphi_{2,\bm{k}, \alpha'}, \varphi_{3,\bm{k}, \alpha'})^{\operatorname{T}}$, the inter-band velocity at the $X$-point is $V^{\alpha'\alpha}_{x, \bm{k}} |_{\bm{k}_X} = \left.\mathcal{D}_{\bm{k}, \alpha'}^{\dagger} \partial_{k_x} H_{\text{f}} \mathcal{F}_{\bm{k}, \alpha} \right|_{\bm{k}_X} = \left. \left( \varphi_{1,\bm{k}}^{\dagger} \partial_{k_x} S_{1,\bm{k}} \phi_{1,\bm{k}} + \varphi_{1,\bm{k}}^{\dagger} \partial_{k_x} S_{2,\bm{k}} \phi_{2,\bm{k}}\right) \right| _{\bm{k}_X} \sim a\operatorname{max}(\mathcal{E}_{1,\Gamma})$, which reflects the largest hopping amplitude in the system (assumed to be along $x$). This implies $ V^{\alpha'\alpha}_{x, \bm{k}} |_{\bm{k}_X} \gg \Delta_X$, resulting in a large non-Abelian quantum metric at the $X$-point:
\begin{equation}
    \mathcal{G}^{\alpha\alpha}_{xx}({\bm{k}}_X) \sim \sum_{\alpha'} \left. \frac{(V^{\alpha'\alpha}_{x, \bm{k}})^{\dagger} (V^{\alpha'\alpha}_{x, \bm{k}})}{(\Delta_{\bm{k}})^2} \right|_{\bm{k}_X} \gg a^2.
\end{equation}
%Notably, reducing the energy scales in $S_{2,X}$ while holding Eqs.(\ref{Eq_Local_Atomic_Constraint}) and (\ref{Eq_nontrivialize_FB_condition})
%further enhances the $X$-point quantum metric due to the decreased band gap. 
And it can be further enhanced by reducing the $X$-point band gap.
Therefore, a hot spot region near the $X$-point is expected to emerge in the BZ distribution of the quantum metric.
%In contrast, with the constraint in Eq.~(\ref{Eq_Local_Atomic_Constraint}), the $\Gamma$-point inter-band velocity would be $V^{\alpha'\alpha}_{x, \bm{k}} |_{\bm{k}_{\Gamma}} = \left. \mathcal{D}_{\bm{k}, \alpha'}^{\dagger} \partial_{k_x} H_{\text{f}} \mathcal{F}_{\bm{k}, \alpha} \right|_{\bm{k}_{\Gamma}} \approx \left. \varphi_{1,\bm{k}}^{\dagger} \partial_{k_x} S_{2,\bm{k}} \phi_{2,\bm{k}} \right| _{\bm{k}_{\Gamma}} \sim a\operatorname{max}(\mathcal{E}_{2,X}) \sim \Delta_X$. 

\begin{comment}
Since the $X$-point band gap corresponds to the smallest singular value of $S_{\bm{k}}$, it is much smaller than the largest energy scale of the system: $\Delta_{X} \sim \operatorname{max}(\mathcal{E}_{2,X}) \ll \operatorname{max}(\mathcal{E}_{1,\Gamma})$. Meanwhile, in tight-binding models one has $\bra{\mathcal{D}_{\bm{k}}} \partial_{k_x} H \ket{\mathcal{F}_{\bm{k}}} |_{\bm{k}_X} \sim a \operatorname{max}(\mathcal{E}_{1,\Gamma})$, reflecting the largest hopping amplitude of the system, assumed along the $x$-direction.
Consequently, the $xx$-component of the quantum metric for the flat bands, significantly exceeds the lattice scale:
\begin{equation}
    \mathcal{G}_{xx}({\bm{k}}_X) \sim \left. \frac{(\partial_{k_x} H)(\partial_{k_x} H)}{(\Delta)^2} \right|_{\bm{k}_X} \gg a^2.
\end{equation}
\end{comment}

In the variant Lieb lattice, for example, the energy scales for $S_{1,\bm{k}}$ and $S_{2,\bm{k}}$ are $\mathcal{E}_{1,\Gamma} = \{J\}$, $\mathcal{E}_{2,\Gamma} = \{\alpha \delta J\}$, yielding $\lambda_{\Gamma} = \alpha\delta \ll 1$, which satisfies the local atomic constraint in Eq.~\eqref{Eq_Local_Atomic_Constraint}. Meanwhile, it hosts a direct band gap at the X-point, where the energy scales are $\mathcal{E}_{1,X} = \{\delta J\}$, $\mathcal{E}_{2,X} = \{\alpha \delta J\}$. For $\alpha \sim 1$, these scales are comparable, satisfying Eq.~(\ref{Eq_nontrivialize_FB_condition}) and resulting in a band gap $\Delta_{X} \sim \delta J$. Consequently, the variant Lieb lattices feature a peak of the quantum metric distribution near the X-point which can be further enhanced by reducing the band gap $\Delta_{X}$.

\section{Acknowledgment}
 The authors would like to thank Xun-jiang Luo, Zi-Ting Sun and Hua Jiang for inspiring discussions. K. T. L. acknowledges the support of the Ministry of Science and Technology, China, and Hong Kong Research Grant Council through Grants No. 2020YFA0309600, RFS2021-6S03, C6025-19G, C6053-23G, AoE/P-701/20, 16310520, 16310219, 16307622, 16311424 and 16309223.

\bibliography{QMLTopologyNotes}
%\end{document}    % no \documentclass or \begin{document} inside Main.tex
		
		% --- Supplementary section ---
		\clearpage	
		% Reset counters and redefine numbering
		\setcounter{equation}{0}
		\setcounter{section}{0}
		\setcounter{figure}{0}
		\setcounter{table}{0}
		\setcounter{page}{1}
		\renewcommand{\theequation}{S\arabic{equation}}
		\renewcommand{\thesection}{\Roman{section}}
		\renewcommand{\thefigure}{S\arabic{figure}}
		\renewcommand{\thetable}{\arabic{table}}
		\renewcommand{\tablename}{Supplementary Table}
		\renewcommand{\bibnumfmt}[1]{[S#1]}
		\renewcommand{\citenumfont}[1]{S#1}

		% Switch to one column
		\onecolumngrid
		
		\clearpage
%		\begin{center}
%			\textbf{\large Supplementary Note for ``Universal Boundary-Mode Localization from Quantum Metric Length''}\\[0.5em]
%			Xinglei Ma, Jin-Xin Hu\footnote{jhuphy@ust.hk}, K. T. Law\footnote{phlaw@ust.hk}\\[0.5em]
%			Department of Physics, Hong Kong University of Science and Technology, Clear Water Bay, Hong Kong, China\\[0.5em]
%			\today
%		\end{center}

		\begin{center}
			\textbf{\large Supplementary Note for "Non-Abelian Quantum Metric Governed Topological Boundary-Mode Localization"}
			\\[0.5em]
			Xing-Lei Ma\textsuperscript{1}, Jin-Xin Hu\textsuperscript{1}\textsuperscript{*}, K. T. Law\textsuperscript{1}\textsuperscript{\dag} \\[0.5em]
			{\footnotesize\itshape
				\textsuperscript{1}Department of Physics, Hong Kong University of Science and Technology, Clear Water Bay, Hong Kong, China\\
			}
			{\footnotesize\textnormal
				\textsuperscript{*}jhuphy@ust.hk \quad \textsuperscript{\dag}phlaw@ust.hk
			}
		\end{center}

		\tableofcontents
		\vspace{2em}

\section{\label{Sec_Quantum_Metric}Supplementary Note I: Abelian and non-Abelian quantum metric tensor}

We start by revisiting the quantum geometric tensor for a given Hamiltonian $H(\bm{\kappa})$ living in a $d$-dimensional parameter space $\mathcal{K}$, with coordinates $\bm{\kappa} = (\kappa_1, \kappa_2, ... \kappa_d)$. Let us take an eigenspace $\mathcal{H}_{E}$ from the $h$-dimensional Hilbert space of $H(\bm{\kappa})$ with eigenvalue $E(\bm{\kappa})$ and degeneracy $g$. Suppose $\mathcal{H}_{E}$ is well isolated; i.e., $E(\bm{\kappa})$ remains separated from other eigenvalues across the parameter space. Provided a complete orthonormal basis $\mathcal{B} = \{\psi_1(\bm{\kappa}),... \psi_g(\bm{\kappa})\}$ in $\mathcal{H}_{E}$, any arbitrary state in the eigenspace can be expressed as
\begin{equation}
	\Psi_{0,\bm{\kappa}} =  \sum^g_{{\alpha}} C_{\alpha}(\bm{\kappa}) \psi_{\alpha}(\bm{\kappa}).
	\label{Eq_Psi_kappa}
\end{equation}
Under the basis set $\mathcal{B}$, the state can also be represented as a vector $\mathcal{C}_{\bm{\kappa}} = [C_1(\bm{\kappa}), ...  C_g(\bm{\kappa})]^{\operatorname{T}}$, and any two states can be related by a $U(g)$ local gauge transformation, rendering $\mathcal{H}_{E}$ a $U(g)$ vector bundle over $\mathcal{K}$. As a section in the vector bundle, $\Psi_{0,\bm{\kappa}}$ can be further equipped with intrinsic structures, one of which is the metric. To define it, consider the distance $D(\bm{\kappa}) $ between two neighboring states on $\mathcal{K}$:
\begin{equation}
	\begin{aligned}
		D_{\Psi_0}(\bm{\kappa}) 
		&= 1 - \left|\Psi_{0,\bm{\kappa}}^{\dagger} \Psi_{0,\bm{\kappa}+ \diff \bm{\kappa}} \right|^2 \\
		%&= \sum_{ab} \operatorname{Tr}(\partial_a P_0(\bm{\kappa})\partial_b P_0(\bm{\kappa})) \diff \kappa_a \diff \kappa_b \\
		&= \sum^d_{ab} \Re \Big[ \partial_{a} \Psi_{0,\bm{\kappa}}^{\dagger} \left(1 - \Psi_{0,\bm{\kappa}} \Psi_{0,\bm{\kappa}}^{\dagger}\right)    \partial_{b} \Psi_{0,\bm{\kappa}} \Big] \diff \kappa_a \diff \kappa_b  \\
		%&= \sum^d_{ab} \Re \left[\mathcal{Q}^{0}_{ab}(\bm{\kappa}) \right]  \diff \kappa_a \diff \kappa_b. \\
		&= \sum^d_{ab} \mathcal{G}^0_{ab}(\bm{\kappa})   \diff \kappa_a \diff \kappa_b, \\
	\end{aligned}
	\label{Eq_distance_of2_vectors}
\end{equation}
where $\partial_{a} \coloneqq \partial/\partial_{\kappa_a}$.
Here, $\mathcal{G}^0_{ab}(\bm{\kappa})$ is the metric tensor for the state $\Psi_{0,\bm{\kappa}}$ on $\mathcal{K}$. In particular, when $\mathcal{H}_{E}$ is non-degenerate, i.e., $g=1$, $\mathcal{G}^0_{ab}(\bm{\kappa})$ recovers the Abelian quantum metric tensor in the original basis $\mathcal{B}$:
\begin{equation}
	\begin{aligned}
		\mathcal{G}_{ab}(\bm{\kappa}) = \Re \left[ \partial_{a} \psi_1^{\dagger}(\bm{\kappa}) \left(1 - \psi_1(\bm{\kappa}) \psi^{\dagger}_1(\bm{\kappa})\right)   \partial_{b} \psi_1(\bm{\kappa}) \right].
	\end{aligned}
	\label{Eq_Abelian_QM_tensor}
\end{equation}

When $g>1$, the metric tensor $\mathcal{G}^0_{ab}(\bm{\kappa})$ can then be related to the non-Abelian quantum metric $\mathcal{G}^{\alpha\alpha'}_{ab}(\bm{\kappa})$ in the degenerate manifold by:
\begin{equation}
	\begin{aligned}
		\mathcal{G}^0_{ab}(\bm{\kappa}) 
		&= \sum^{g}_{\alpha \alpha'}  \mathcal{C}_{\bm{\kappa},\alpha}^{\dagger} \mathcal{G}^{\alpha\alpha'}_{ab}(\bm{\kappa}) \mathcal{C}_{\bm{\kappa},\alpha'} \\
		&\quad  + \sum^{g-1}_{0_{\perp}\in \mathcal{H}_E} \Re \left( \partial_a\Psi_{0_{\perp},\bm{\kappa}}^{\dagger} \Psi_{0,\bm{\kappa}} \Psi_{0,\bm{\kappa}}^{\dagger} \partial_b\Psi_{0_{\perp},\bm{\kappa}} \right),
	\end{aligned}
	\label{Eq_Relation_Abelian_and_nonAbelian_QM}
\end{equation}
where $\Psi_{0_{\perp},\bm{\kappa}}$ denotes one of the $g-1$ states in $\mathcal{H}_{E}$ that is perpendicular to $\Psi_{0,\bm{\kappa}}$. Note that the second term is non-vanishing unless considering the quantum adiabatic limit \cite{PhysRevA.90.022104}, and this has not been clarified in \cite{PhysRevA.109.043305, PhysRevA.105.012210}. To see this, we decompose $\partial_a \Psi_{0,\bm{\kappa}} = D_a \Psi_{0,\bm{\kappa}} + \sum^{g}_{\alpha} C_{\alpha}(\bm{\kappa}) \left[1-P_0(\bm{\kappa})\right] \partial_a \psi_{\alpha}(\bm{\kappa})$, with $D_a \Psi_{0,\bm{\kappa}} = \sum^{g}_{\alpha} \left[ \partial_a C_{\alpha}(\bm{\kappa})\right] \psi_{\alpha}(\bm{\kappa})  + \sum^{g}_{\alpha} C_{\alpha}(\kappa) P_0(\bm{\kappa}) \partial_a \psi_{\alpha}(\bm{\kappa})$ where the second term is related to the Christoffel symbol in~\cite{hetenyi2023fluctuations}, and the projector $P_0(\bm{\kappa}) \coloneqq \sum_{\alpha}^g \psi_{\alpha}(\bm{\kappa}) \psi^{\dagger}_{\alpha}(\bm{\kappa})$. If $\Psi_{0,\bm{\kappa}}$ evolves in the quantum adiabatic limit in the parameter space with~\cite{PhysRevB.81.245129}
\begin{equation}
	\begin{aligned}
		D_a \Psi_{0,\bm{\kappa}} = 0,
	\end{aligned}
	\label{Eq_quantum_adiabatic_condition}
\end{equation}
the second term in Eq.~(\ref{Eq_Relation_Abelian_and_nonAbelian_QM}) will vanish and lead to $\mathcal{G}^0_{ab}(\bm{\kappa}) = 
\sum^{g}_{\alpha \alpha'}  \mathcal{C}_{\bm{\kappa},\alpha}^{\dagger} \mathcal{G}^{\alpha\alpha'}_{ab}(\bm{\kappa}) \mathcal{C}_{\bm{\kappa},\alpha'}$. The non-Abelian quantum metric tensor $\mathcal{G}^{\alpha\alpha'}_{ab}(\bm{\kappa})$ is defined as follows:
\begin{equation}
	\begin{aligned}
		\mathcal{G}^{\alpha\alpha'}_{ab}(\bm{\kappa})
		&\equiv  \Re  \left[  \partial_a \psi_{\alpha}^{\dagger} \left( 1 - \sum^{g}_{\beta \in \mathcal{H}_E} \psi_{\beta}\psi_{\beta}^{\dagger} \right) \partial_b \psi_{\alpha'}  \right] \\
		&= \sum^{h-g}_{\beta \notin \mathcal{H}_E}  \Re \left[ \frac{\psi_{\alpha}^{\dagger} \partial_a H(\bm{\kappa}) \psi_{\beta} \psi_{\beta}^{\dagger} \partial_b H(\bm{\kappa}) \psi_{\alpha'}}{(E(\bm{\kappa}) - E_{\beta}(\bm{\kappa}))^2}   \right],
	\end{aligned}
	\label{Eq_non_Abelian_QM}
\end{equation}
which, together with the non-Abelian Berry curvature $\mathrm{\Omega}_{ab}(\bm{\kappa})$, constitutes the non-Abelian quantum geometric tensor,
\begin{equation}
	\begin{aligned}
		\mathcal{Q}^{\alpha\alpha'}_{ab}(\bm{\kappa})  = \mathcal{G}^{\alpha\alpha'}_{ab}(\bm{\kappa}) + \frac{1}{2i} \mathrm{\Omega}^{\alpha\alpha'}_{ab}(\bm{\kappa}).
	\end{aligned}
	\label{Eq_Quantum_geometric_tensor}
\end{equation}
In 2D, both the real part and imaginary part are nontrivial, and their interesting interplay is the focus of this work.
Numerically we use the second line in Eq.~(\ref{Eq_non_Abelian_QM}) for calculation to avoid the gauge problem. The proof for Eq.~(\ref{Eq_Relation_Abelian_and_nonAbelian_QM}) and Eq.~(\ref{Eq_non_Abelian_QM}) is given in Appendix~\ref{appendix_nonAbelian_QM}.

\subsection{Proof for Eqs.~(\ref{Eq_Relation_Abelian_and_nonAbelian_QM}) and (\ref{Eq_non_Abelian_QM}).}
\label{appendix_nonAbelian_QM}

Starting from the Hamiltonian $H(\bm{\kappa})$ with $g$-fold degenerate eigenspace $\mathcal{H}_E$, we decompose it with three sets of projectors: $P_{0, \bm{\kappa}} \equiv \Psi_{0,\bm{\kappa}} \Psi_{0,\bm{\kappa}}^{\dagger}$ for the target state; $P_{0_{\perp}, \bm{\kappa}} \equiv \sum^{g -1}_{0_{\perp} \in S_{\mathcal{F}} }  \Psi_{0_{\perp}, \bm{\kappa}} \Psi_{0_{\perp}, \bm{\kappa}}^{\dagger}$ for the states in $\mathcal{H}_E$ orthogonal to $\Psi_{0, \bm{\kappa}}$; and $Q_{0, \bm{\kappa}} \equiv 1-(P_{0, \bm{\kappa}}+P_{0_{\perp}, \bm{\kappa}}) = \sum^{h -g}_{\beta \notin S_{\mathcal{F}}} \Psi_{\beta, \bm{\kappa}} \Psi_{\beta,\bm{\kappa}}^{\dagger}$for states outside $\mathcal{H}_E$, satisfying $H(\bm{\kappa}) Q_{0, \bm{\kappa}} = \sum^{h -g}_{\beta \notin S_{\mathcal{F}}} E_{\beta}(\kappa) \Psi_{\beta, \bm{\kappa}} \Psi_{\beta,\bm{\kappa}}^{\dagger}$. The Hamiltonian can then be expressed as:
\begin{equation}
	\begin{aligned}
		H(\bm{\kappa}) = E(\kappa) P_{0, \bm{\kappa}} + E(\kappa) P_{0_{\perp}, \bm{\kappa} }
		+ Q_{0, \bm{\kappa} } H(\bm{\kappa} ) Q_{0, \bm{\kappa} }.
	\end{aligned}
\end{equation}
Consider
\begin{equation}
	\begin{aligned}
		\partial_{a} H(\bm{\kappa} ) P_{0, \bm{\kappa}} &= \partial_{a} E(\kappa) P_{0, \bm{\kappa}}  + E(\kappa) \partial_{a}P_{0, \bm{\kappa}} P_{0, \bm{\kappa}} + E(\kappa) \partial_{a}P_{0_{\perp}, \bm{\kappa} } P_{0, \bm{\kappa}} 
		- Q_{0, \bm{\kappa} } H(\bm{\kappa} ) (\partial_{a}P_{0, \bm{\kappa}} + \partial_{a}P_{0_{\perp}, \bm{\kappa} }) P_{0, \bm{\kappa}}\\ 
		&= \partial_{a} E(\kappa) P_{0, \bm{\kappa}} + E(\kappa) \partial_{a}P_{0, \bm{\kappa}} P_{0, \bm{\kappa}} + E(\kappa) \partial_{a}P_{0_{\perp}, \bm{\kappa} } P_{0, \bm{\kappa}} 
		- H(\bm{\kappa} ) (\partial_{a}P_{0, \bm{\kappa}} + \partial_{a}P_{0_{\perp}, \bm{\kappa} }) P_{0, \bm{\kappa}} \\ & \ \ \ + E(\kappa) (P_{0, \bm{\kappa}}+P_{0_{\perp}, \bm{\kappa} }) \partial_{a}(P_{0, \bm{\kappa}} + P_{0_{\perp}, \bm{\kappa} }) P_{0, \bm{\kappa}}. \\
	\end{aligned}
	\label{Eq_partialHP0}
\end{equation}
For any two projectors $P_{1,\bm{\kappa} }$ and $P_{2,\bm{\kappa} }$, we observe that
\begin{equation}
	\begin{aligned} 
		P_{1,\bm{\kappa} } \partial_{a} P_{1,\bm{\kappa} } P_{2,\bm{\kappa} } &= P_{1,\bm{\kappa} } \left(\partial_{a} P_{1,\bm{\kappa} }  P_{1,\bm{\kappa} } + P_{1,\bm{\kappa} }\partial_{a} P_{1,\bm{\kappa} }\right ) P_{2,\bm{\kappa} }   \\
		&= P_{1,\bm{\kappa} }\partial_{a} P_{1,\bm{\kappa} }  P_{1,\bm{\kappa} } P_{2,\bm{\kappa} }  + P_{1,\bm{\kappa} }\partial_{a} P_{1,\bm{\kappa} } P_{2,\bm{\kappa} }, 
	\end{aligned}
\end{equation}
implying that $P_{1,\bm{\kappa} }\partial_{a} P_{1,\bm{\kappa} }  P_{1,\bm{\kappa} } P_{2,\bm{\kappa} } = 0$. In the specific case where $P_{1,\bm{\kappa} } = (P_{0, \bm{\kappa}} + P_{0_{\perp}, \bm{\kappa} })$ and $P_{2,\bm{\kappa} } = P_{0, \bm{\kappa}}$, one has $ P_{1,\bm{\kappa} } P_{2,\bm{\kappa} } = P_{2,\bm{\kappa} }$, which leads to $P_{1,\bm{\kappa} }\partial_{a} P_{1,\bm{\kappa} } P_{2,\bm{\kappa} } = 0$. As a result, the fourth term in the last step of Eq.~(\ref{Eq_partialHP0}) vanishes, and Eq.~(\ref{Eq_partialHP0}) then simplifies to:
\begin{equation}
	\begin{aligned} 
		\partial_{a} P_{0,\bm{\kappa} }  P_{0,\bm{\kappa} } =  -Q_{0,\bm{\kappa} } \frac{1}{H(\bm{\kappa} ) - E(\kappa)} Q_{0,\bm{\kappa} } \partial_{a} H(\bm{\kappa} ) P_{0,\bm{\kappa} } - \partial_{a}  P_{0_{\perp},\bm{\kappa} } P_{0,\bm{\kappa} }.
	\end{aligned}
	\label{Eq_partialP0P01}
\end{equation}
Similarly,
\begin{equation}
	\begin{aligned} 
		P_{0,\bm{\kappa} } \partial_{a} P_{0,\bm{\kappa} } =  - P_{0,\bm{\kappa} } \partial_{a} H(\bm{\kappa} )  Q_{0,\bm{\kappa} } \frac{1}{H(\bm{\kappa} ) - E(\kappa)} Q_{0,\bm{\kappa} } - P_{0,\bm{\kappa} } \partial_{a}  P_{0_{\perp},\bm{\kappa} }.
	\end{aligned}
	\label{Eq_partialP0P02}
\end{equation}

Combining Eq.~(\ref{Eq_partialP0P01}) and Eq.~(\ref{Eq_partialP0P02}), it follows that
\begin{equation}
	\begin{aligned} 
		\partial_{a} P_{0,\bm{\kappa} } =  - \left[ Q_{0,\bm{\kappa} } \frac{1}{H(\bm{\kappa} ) - E(\kappa)} Q_{0,\bm{\kappa} } \partial_{a} H(\bm{\kappa} ) P_{0,\bm{\kappa} } + P_{0,\bm{\kappa} } \partial_{a} H(\bm{\kappa} )  Q_{0,\bm{\kappa} } \frac{1}{H(\bm{\kappa} ) - E(\kappa)} Q_{0,\bm{\kappa} } + (\partial_{a}  P_{0_{\perp},\bm{\kappa} } P_{0,\bm{\kappa} } + P_{0,\bm{\kappa} } \partial_{a}  P_{0_{\perp},\bm{\kappa} })
		\right].
	\end{aligned}
\end{equation}

Then, consider the operator trace for $\partial_{a} P_{0,\bm{\kappa} }\partial_{b} P_{0,\bm{\kappa} }$:
\begin{equation}
	\begin{aligned} 
		&\operatorname{Tr} \left( \partial_{a} P_{0,\bm{\kappa} }\partial_{b} P_{0,\bm{\kappa} } \right) \\
		=  & \operatorname{Tr} \left\{ \left[ Q_{0,\bm{\kappa} } \frac{1}{H(\bm{\kappa} ) - E(\kappa)} Q_{0,\bm{\kappa} } \partial_{a} H(\bm{\kappa} ) P_{0,\bm{\kappa} } + P_{0,\bm{\kappa} } \partial_{a} H(\bm{\kappa} )  Q_{0,\bm{\kappa} } \frac{1}{H(\bm{\kappa} ) - E(\kappa)} Q_{0,\bm{\kappa} } + (\partial_{a}  P_{0_{\perp},\bm{\kappa} } P_{0,\bm{\kappa} } + P_{0,\bm{\kappa} } \partial_{a}  P_{0_{\perp},\bm{\kappa} }) 
		\right] \right.  \\
		& \ \ \ \ \cdot \left. \left[ Q_{0,\bm{\kappa} } \frac{1}{H(\bm{\kappa} ) - E(\kappa)} Q_{0,\bm{\kappa} } \partial_{b} H(\bm{\kappa} ) P_{0,\bm{\kappa} } + P_{0,\bm{\kappa} } \partial_{b} H(\bm{\kappa} )  Q_{0,\bm{\kappa} } \frac{1}{H(\bm{\kappa} ) - E(\kappa)} Q_{0,\bm{\kappa} } + (\partial_{b}  P_{0_{\perp},\bm{\kappa} } P_{0,\bm{\kappa} } + P_{0,\bm{\kappa} } \partial_{b}  P_{0_{\perp},\bm{\kappa} }) 
		\right] \right\} \\
		= &  \sum_{\beta \notin \mathcal{H}_E}^{h-g} \left[ \frac{\Psi_0^{\dagger} \partial_{a} H(\bm{\kappa} ) \Psi_{\beta} \Psi_{\beta}^{\dagger} \partial_{b} H(\bm{\kappa} ) P_{0,\bm{\kappa} }}{\left(E_{\beta}(\bm{\kappa} ) - E(\kappa)\right)^2}  + (a \leftrightarrow b) \right]  
		+  \operatorname{Tr} \left( \partial_{a} P_{0_{\perp},\bm{\kappa} } P_{0,\bm{\kappa} } \partial_{b} P_{0_{\perp},\bm{\kappa} } P_{0,\bm{\kappa} } \right.  \\
		& + \left. \partial_{a} P_{0_{\perp},\bm{\kappa} } P_{0,\bm{\kappa} } \partial_{b} P_{0_{\perp},\bm{\kappa} } + P_{0,\bm{\kappa} } \partial_{a} P_{0_{\perp},\bm{\kappa} } \partial_{b} P_{0_{\perp},\bm{\kappa} } P_{0,\bm{\kappa} } + P_{0,\bm{\kappa} } \partial_{a} P_{0_{\perp},\bm{\kappa} } P_{0,\bm{\kappa} } \partial_{b} P_{0_{\perp},\bm{\kappa} } \right)  \\
		&  + \sum_{\beta \notin \mathcal{H}_E}^{h-g} \operatorname{Tr} \left[ \frac{P_{\beta, \bm{\kappa} }}{E_{\beta}(\bm{\kappa} ) - E(\kappa)}  \partial_{a}H(\bm{\kappa} ) P_{0,\bm{\kappa} } \partial_{b} P_{0_{\perp},\bm{\kappa} }    +   \partial_{a}  P_{0_{\perp},\bm{\kappa} } P_{0,\bm{\kappa} } \partial_{b}H(\bm{\kappa} ) \frac{P_{\beta, \bm{\kappa} }}{E_{\beta}(\bm{\kappa} ) - E(\kappa)}  + (a \leftrightarrow b) \right]  ,
	\end{aligned}
	\label{Eq_Trace_ab}
\end{equation}
where, in the second step, we have used:
\begin{equation}
	\begin{aligned} 
		Q_{0,\bm{\kappa} } \frac{1}{H(\bm{\kappa} ) - E(\kappa)} Q_{0,\bm{\kappa} } = \sum_{\beta \notin S_{\mathcal{F}}}^{h-g} \frac{1}{E_{\beta}(\bm{\kappa} ) - E(\kappa)} \Psi_{\beta}(\bm{\kappa} ) \Psi_{\beta}^{\dagger}(\bm{\kappa} ).
	\end{aligned}
\end{equation}

Note
\begin{equation}
	\begin{aligned} 
		P_{0,\bm{\kappa} } \partial_{a}  P_{0_{\perp},\bm{\kappa} }  &= P_{0,\bm{\kappa} } \partial_{a}  P_{0_{\perp},\bm{\kappa} } P_{0_{\perp},\bm{\kappa} }, \\
		\partial_{a}  P_{0_{\perp},\bm{\kappa} } P_{0,\bm{\kappa} } &=  P_{0_{\perp},\bm{\kappa} } \partial_{a}  P_{0_{\perp},\bm{\kappa} } P_{0,\bm{\kappa} } , \\
	\end{aligned}
\end{equation}
and
\begin{equation}
	\begin{aligned} 
		P_{0,\bm{\kappa} } \partial_{a}  P_{0_{\perp},\bm{\kappa} } &= \partial_{a} (P_{0,\bm{\kappa} } P_{0_{\perp},\bm{\kappa} }) - \partial_{a}P_{0,\bm{\kappa} } P_{0_{\perp},\bm{\kappa} }  = - \partial_{a}P_{0,\bm{\kappa} } P_{0_{\perp},\bm{\kappa} }.
	\end{aligned}
\end{equation}

Then Eq.~(\ref{Eq_Trace_ab}) can be further simplified:
\begin{equation}
	\begin{aligned} 
		\operatorname{Tr} \left( P_{0,\bm{\kappa} }\partial_{b} P_{0,\bm{\kappa} } \right) &=  \sum_{\beta \notin \mathcal{H}_E}^{h-g} \left[ \frac{\Psi_0^{\dagger} \partial_{a} H(\bm{\kappa} ) \Psi_{\beta} \Psi_{\beta}^{\dagger} \partial_{b} H(\bm{\kappa} ) P_{0,\bm{\kappa} }}{\left(E_{\beta}(\bm{\kappa} ) - E(\kappa)\right)^2}  + (a \leftrightarrow b) \right]  + \operatorname{Tr} \left(  \partial_{a} P_{0_{\perp},\bm{\kappa} } P_{0,\bm{\kappa} } \partial_{b} P_{0_{\perp},\bm{\kappa} } + \partial_{a}  P_{0,\bm{\kappa} } P_{0_{\perp},\bm{\kappa} } \partial_{b} P_{0,\bm{\kappa} } \right) \\
		&= \sum_{\beta \notin \mathcal{H}_E}^{h-g} \left[ \frac{\Psi_0^{\dagger} \partial_{a} H(\bm{\kappa} ) \Psi_{\beta} \Psi_{\beta}^{\dagger} \partial_{b} H(\bm{\kappa} ) P_{0,\bm{\kappa} }}{\left(E_{\beta}(\bm{\kappa} ) - E(\kappa)\right)^2}  + (a \leftrightarrow b) \right]  + \sum_{0_{\perp} \in \mathcal{H}_E}^{g-1} 2\Re \left( \partial_a\Psi_{0_{\perp},\bm{\kappa} }^{\dagger} \Psi_{0,\bm{\kappa} } \Psi_{0,\bm{\kappa} }^{\dagger} \partial_b\Psi_{0_{\perp},\bm{\kappa} } \right). 
	\end{aligned}
	\label{Eq_Trace_ab2}
\end{equation}

In the above procedures we have repeatedly applied the following useful relation:
\begin{equation}
	\begin{aligned}
		\partial_{a} P_{0, \bm{\kappa}} = \partial_{a} (P_{0, \bm{\kappa}}P_{0, \bm{\kappa}}) = \partial_{a} P_{0, \bm{\kappa}} P_{0, \bm{\kappa}} + P_{0, \bm{\kappa}} \partial_{a} P_{0, \bm{\kappa}}. \\
		% &P_{0, \bm{\kappa}}	\partial_{a} P_{0, \bm{\kappa}} P_{0, \bm{\kappa}} = 0 
	\end{aligned}
\end{equation}

With Eq.~(\ref{Eq_Trace_ab2}), let us re-examine the quantum metric tensor of the states $\Psi_{0,\bm{\kappa}}$ on the degenerate eigenspace $\mathcal{H}_E$, defined in Eq.~(\ref{Eq_distance_of2_vectors}):
\begin{equation}
	\begin{aligned}
		\mathcal{G}^0_{ab}(\bm{\kappa})
		&= \Re \left[ \partial_{a} \Psi_{0,\bm{\kappa}}^{\dagger} \left(1 - \Psi_{0, \bm{\kappa}}  \Psi^{\dagger}_{0,\bm{\kappa}}\right)   \partial_{b} \Psi_{0, \bm{\kappa}}  \right]  \\
		&=\frac{1}{2}  \operatorname{Tr}(\partial_a P_{0,\bm{\kappa}}\partial_b P_{0,\bm{\kappa}})  \\
		& \begin{aligned} =
			\frac{1}{2} 
			\left\{ \sum_{\beta \notin \mathcal{H}_E}^{h-g} \right. \left. \left[ \frac{\Psi_{0,\bm{\kappa}}^{\dagger} \partial_{a} H(\bm{\kappa} ) \Psi_{\beta, \bm{\kappa}} \Psi_{\beta, \bm{\kappa}}^{\dagger} \partial_{b} H(\bm{\kappa} ) \Psi_{0,\bm{\kappa} }}{\left(E_{\beta}(\bm{\kappa} ) - E(\kappa)\right)^2}  + (a \leftrightarrow b) \right]  \right.
			+ \left. \sum_{0_{\perp} \in \mathcal{H}_E}^{g-1} 2\Re \left( \partial_a\Psi_{0_{\perp},\bm{\kappa} }^{\dagger} \Psi_{0,\bm{\kappa} } \Psi_{0,\bm{\kappa} }^{\dagger} \partial_b\Psi_{0_{\perp},\bm{\kappa} } \right)   \right\} 
		\end{aligned} \\
		&= \left[\sum^{g}_{\alpha \alpha'}  \mathcal{C}_{\bm{\kappa},\alpha}^{\dagger} \mathcal{G}^{\alpha\alpha'}_{ab}(\bm{\kappa}) \mathcal{C}_{\bm{\kappa},\alpha'} + \sum_{0_{\perp} \in \mathcal{H}_E}^{g-1} \Re \left( \partial_a\Psi_{0_{\perp},\bm{\kappa} }^{\dagger} \Psi_{0,\bm{\kappa} } \Psi_{0,\bm{\kappa} }^{\dagger} \partial_b\Psi_{0_{\perp},\bm{\kappa} } \right)  \right], \\
	\end{aligned}
	\label{Eq_FB_distance_nonAbelian_QM_extra_term}
\end{equation}
where the non-Abelian quantum metric $\mathcal{G}^{\alpha\alpha'}_{ab}(\bm{\kappa} )$ is defined as:
\begin{equation}
	\begin{aligned} 
		\mathcal{G}^{\alpha\alpha'}_{ab} (\bm{\kappa} )
		&=  \Re  \left[  \partial_a \psi_{\alpha}^{\dagger} \left( 1 - \sum^{g}_{\beta} \psi_{\beta}\psi_{\beta}^{\dagger} \right) \partial_b \psi_{\alpha'}  \right] \\
		&= \sum_{\beta \notin \mathcal{H}_E}  \Re \left[ \frac{\psi_{\alpha}^{\dagger} \partial_a H(\bm{\kappa}) \psi_{\beta} \psi_{\beta}^{\dagger} \partial_b H(\bm{\kappa}) \psi_{\alpha'}}{(E_{f} - E_{\beta})^2}   \right]. \\
		%&= \frac{1}{2}   \sum_{\beta \notin \mathcal{H}_E}^{h} \left(\frac{\psi_{\alpha}^{\dagger} \partial_a H(\bm{\kappa} ) \psi_{\beta} \psi_{\beta}^{\dagger} \partial_b H(\bm{\kappa} ) \psi_{\alpha'}}{(E_{f} - E_{\beta})^2} + (a\leftrightarrow b) \right). \\
	\end{aligned}
	\label{Eq_nonAbelian_QM_def}
\end{equation}
This finishes the proof for Eqs.~(\ref{Eq_Relation_Abelian_and_nonAbelian_QM}) and (\ref{Eq_non_Abelian_QM}).

\section{Supplementary Note II: Spatial behaviors of topological boundary modes}

\subsection{Eigen-equations: From full Hamiltonian to effective model} %--Proof for Eq.~(\ref{Eq_edge_states_eigen_equation}).}
\label{appendix_OBC_effective_Hamiltonian}

\begin{comment}
\subsection{Wave functions for TBM}
To reveal the spatial behaviors for the boundary modes, let us return to the effective model obtained by projecting the full model onto the degenerate manifolds. In the topological regime with open boundary conditions (OBC), flat band spectrum reveals in-gap TBMs, as shown in Fig.~\ref{Fig_Lieb_QWZ_OBC_spectrum_wavefunc_QSQM}(a, b) for \( x \)-OBC and (e, f) for \( y \)-OBC of the Lieb-QWZ model. Take \( x \)-OBC as an example, $\tilde{\bm{k}} = (k_y,...,k_d)$ remains a good quantum number, we transform $k_x \rightarrow -i\partial_{R_x}$ yielding $H^{\text{eff}}(-i\partial_{R_x}, \tilde{\bm{k}})$, with hybrid Wannier orbitals as bases:
\begin{equation}
    \mathcal{W}_{\alpha}(x-R_x,\tilde{\bm{k}}) = \frac{1}{\sqrt{N}} \sum_{k_x} e^{ik_x (x-R_x)} \mathcal{F}_{\bm{k},\alpha}
    \label{Eq_def_Wannier_function},
\end{equation}
where $R_x$ denotes the lattice translation vector in the \(x\)-direction, and $\mathcal{F}_{\bm{k},\alpha}$ are Bloch vectors of the original exact flat bands.
\end{comment}

With open boundary conditions (OBC) or a domain wall, e.g., along the $x$-direction, $\tilde{\bm{k}} = (k_y,...,k_d)$ remains a good quantum number. The Hamiltonian can be separated into a $\bm{k}$-dependent part $H_{\mathrm{t}}(\bm{k})$ and a spatially varying part $U(x)$, resulting in the following eigen-equation of the topological boundary modes (TBMs):
\begin{equation}
	\begin{aligned}
		\left[H_{\mathrm{t}}(\bm{k}) + U(x)\right] \Psi^{\text{B}}(x, \tilde{\bm{k}}) = E^{\text{B}} \Psi^{\text{B}}(x, \tilde{\bm{k}}).
	\end{aligned}
	\label{Eq_edge_state_eigen_equation}
\end{equation}
To project the full Hamiltonian onto the flat band manifold, we express the TBMs wave functions in terms of the hybrid Wannier orbitals of the flat bands:
\begin{equation}
	\Psi^{\text{B}}(x, \tilde{\bm{k}}) = \sum^{N_f}_{\alpha} \sum_{R_x'} \mathcal{U}_{\alpha}(R_x') \mathcal{W}_{\alpha}(x-R_x', \tilde{\bm{k}}),
	\label{Eq_edge_state_wavefunc_appendix}
\end{equation}
with
\begin{equation}
    \mathcal{W}_{\alpha}(x-R_x,\tilde{\bm{k}}) \equiv \frac{1}{\sqrt{N}} \sum_{k_x} e^{ik_x (x-R_x)} \mathcal{F}_{\bm{k},\alpha}
    \label{Eq_def_Wannier_function},
\end{equation}
where $R_x$ denotes the lattice translation vector in the \(x\)-direction, and $\mathcal{F}_{\bm{k},\alpha}$ are Bloch vectors of the original exact flat bands. The hybrid Wannier orbitals form orthonormal bases in the flat-band manifold:
\begin{equation}
	\int \mathcal{W}^{\dagger}_{\alpha}(x-R_x,\tilde{\bm{k}}) \mathcal{W}_{\beta}(x-R_x',\tilde{\bm{k}}) \diff x = \delta_{\alpha\beta} \delta_{R_xR_x'}
	\label{Eq_Wannier_function_normalization}.
\end{equation}

We then plug Eq.\eqref{Eq_edge_state_wavefunc_appendix} into Eq.\eqref{Eq_edge_state_eigen_equation}, multiply it by $\mathcal{W}^{\alpha}(x-R_x, \tilde{\bm{k}})$ from the left and then integrate over $x$, which results in the following eigen-equation:
\begin{equation}
	\begin{aligned}
		\sum_{\alpha} \sum_{R_x} \left[\mathscr{H}^{\alpha\beta}_{tf}(R_x' - R_x, \tilde{\bm{k}}) - E^{\text{B}} \delta_{R_x R_x'} \delta^{\alpha\beta}  
		+  U^{\alpha \beta}(R_x', R_x)
		\right] \mathcal{U}_{\alpha}(R_x) = 0,
	\end{aligned}
	\label{Eq_edge_state_eigen_equation_Wannier2}
\end{equation}
where we have used:
\begin{equation}
	\begin{aligned}
		&\int \diff x {\mathcal{W}_{\alpha}}^{\dagger}(x-R_x, \tilde{\bm{k}}) H_{\mathrm{t}} \mathcal{W}_{\beta}(x-R_x', \tilde{\bm{k}}) \\
		=& \frac{1}{N_x} \sum_{k_x} e^{ik_x(R_x - R_x')} {\mathcal{F}_{\bm{k}}^{\alpha}}^{\dagger} H_{\mathrm{t}}(\bm{k}) \mathcal{F}_{\bm{k}}^{\beta} \\
		=& \frac{1}{N_x} \sum_{k_x} e^{ik_x(R_x - R_x')} H^{\text{eff}}_{\alpha\beta}(\bm{k}) \\
		\equiv & \mathscr{H}_{\alpha\beta}(R_x' - R_x, \tilde{\bm{k}}),
	\end{aligned}
	\label{Eq_H_eff_Wannier_repre}
\end{equation}
with $\mathscr{H}_{\alpha\beta}(R_x, \tilde{\bm{k}}) \equiv \frac{1}{N_x} \sum_{k_x} e^{-ik_xR_x} H^{\text{eff}}_{\alpha\beta}(\bm{k})$. In the second step of Eq.~\eqref{Eq_H_eff_Wannier_repre}, we have projected the full Hamiltonian onto the original flat-band manifold with $H^{\text{eff}}_{\alpha\beta}(\bm{k}) \equiv {\mathcal{F}_{\bm{k}}^{\alpha}}^{\dagger} H_{\mathrm{t}}(\bm{k}) \mathcal{F}_{\bm{k}}^{\beta}$. We have also defined $U^{\alpha\beta} (R_x, R_x') \equiv \int \mathcal{W}_{\alpha}^{\dagger}(x-R_x, \tilde{\bm{k}}) U(x) \mathcal{W}_{\beta}(x-R_x', \tilde{\bm{k}}) \diff x$ in Eq.~(\ref{Eq_edge_state_eigen_equation_Wannier2}). In general, the projected potential $U^{\alpha\beta} (R_x, R_x')$ depends on the specific Wannier basis one choose. However, notice that for topologically trivial bands, one can choose a gauge for $\mathcal{F}_{\bm{k},\alpha}$ in Eq.~\eqref{Eq_def_Wannier_function} that makes the Bloch states maximally smooth in k-space, producing Wannier orbitals that sharply peak at the orbital site. So effectively the Wannier orbital can be viewed as a delta function with most of its weight concentrated at that peak (despite possessing a long-range tail). Therefore, consider maximally localized wannier orbitals as the basis in Eq.~\eqref{Eq_edge_state_wavefunc_appendix}, the transformed potential approximates to the on-site form of $U(x)$ as:
\begin{equation}
	\begin{aligned}
		U_{\alpha\beta} (R_x, R_x') &\equiv \int {\mathcal{W}_{\alpha}}^{\dagger}(x-R_x, \tilde{\bm{k}}) U(x) \mathcal{W}_{\beta}(x-R_x', \tilde{\bm{k}}) \diff x \\
		&\approx m(x)\Big|_{x=R_x} \delta_{\alpha\beta} \delta_{R_xR_x'} ,\\
	\end{aligned}
	\label{Eq_U_x_transformation_approx}
\end{equation}
where the external potential has been replaced by the spatially varying mass term $m(x)$ controlling the topological regime. %Eq.~\eqref{Eq_U_x_transformation_approx} implies that only by taking maximally localized wannier orbital will the projected potential

Note
\begin{equation}
	\begin{aligned}
		\sum_{R_x'} \mathscr{H}^{\alpha \beta}(R_x'-R_x, \tilde{\bm{k}}) \mathcal{U}_{\alpha}(R_x') &= \sum_{R_x'} \mathscr{H}^{\alpha\beta}(R_x', \tilde{\bm{k}}) \mathcal{U}_{\alpha}(R_x'+R_x) \\
		&= \sum_{R_x'} \mathscr{H}^{\alpha\beta}(R_x', \tilde{\bm{k}}) \mathcal{U}_{\alpha}(x+R_x') \Big|_{x=R_x} \\
		&= \sum_{R_x'} \mathscr{H}^{\alpha\beta}(R_x', \tilde{\bm{k}}) \left[\mathcal{U}_{\alpha}(x) + R_x'\partial_{x}\mathcal{U}_{\alpha}(x)+\frac{1}{2}R_x'^2 \partial^2_x \mathcal{U}_{\alpha}(x) + ... \right] \Big|_{x=R_x} \\
		&= \sum_{R_x'} \mathscr{H}^{\alpha\beta}(R_x', \tilde{\bm{k}}) e^{iR_x'(-i\partial_x)} \mathcal{U}_{\alpha}(x)\Big|_{x=R_x} \\
		&= H^{\text{eff}}_{\alpha \beta} (-i\partial_x, \tilde{\bm{k}}) \mathcal{U}_{\alpha}(x)\Big|_{x=R_x},
	\end{aligned}
\end{equation}
Eq.~(\ref{Eq_edge_state_eigen_equation_Wannier2}) is transformed into the following effective eigen-equation:
\begin{equation}
	\begin{aligned}
		\left[ H_{\alpha \beta}^{\text{eff}}(-i\partial_{R_x}, \tilde{\bm{k}}) + m(R_x)\delta_{\alpha\beta} \right]
		\mathcal{U}^{\beta}(R_x) =
		E^{\text{B}} \mathcal{U}_{\alpha}(R_x),
	\end{aligned}
	\label{Eq_appendix_effective_Hamilt}
\end{equation}
the solution to which gives the modulation vector in the TBMs wave functions in Eq.~\eqref{Eq_edge_state_wavefunc_appendix}.
%This concludes the proof for Eq.~(\ref{Eq_edge_states_eigen_equation}) in the main text. Notice that here the differential operator $\partial_{R_x}$ still acts on discrete lattice.

\subsection{\label{appendix_lowerbound_Topostates} TBMs wave functions}
%{\label{appendix_lowerbound_Topostates} The TBMs wave functions, conventional behavior, geometric and lower bounds for the spread function.}
Consider TBMs arising at a domain wall that separates regions $x<0$ and $x>0$, each characterized by a distinct topological regime. The mass terms on the two sides are: 
\begin{equation}
	m(R_x) = 
	\begin{cases}
		m_l & \text{if $R_x<0$} \\
		m_r & \text{if $R_x>0$},
	\end{cases}
	%\label{}
\end{equation}
with $|m_r| < |\sum_i \tilde{t}_i|$ and $|m_l| > |\sum_i \tilde{t}_i|$. 
%Consider a topological system with open boundary condition so that $m(R_x) = m_r\Theta(R_x) + m_l\Theta(-R_x+1)$
%with $|m_r| < |\sum_i \tilde{t}_i|$ and $m_l \rightarrow \infty$. 
To get an analytical form of the TBMs wave functions, we separate the effective Hamiltonian in Eq.~(\ref{Eq_appendix_effective_Hamilt}) into the quasi-1d part along the OBC direction $H_{1}^{(g)}(\bm{k})$ and the parallel-direction part $H_{d-1}^{(g)}(\bm{k})$:
\begin{equation}
	H^{\alpha \beta}_{\text{eff}}(-i\partial_{R_x}, \tilde{\bm{k}}) + m(R_x) = H_{1}^{(g)}(-i\partial_{R_x},\tilde{\bm{k}}) + H_{d-1}^{(g)}( \tilde{\bm{k}}),
\end{equation}
with
%\begin{equation}
%	\begin{aligned}
	%		H_{1}^{(g)}(\bm{k}) &= -iv_x \gamma_1^{(g)} \partial_{x} + m(x) \gamma_{d+1}^{(g)}  \\
	%		H_{d-1}^{(g)}(\bm{k}) &= \sum_{i=2}^{d} v_i k_i \gamma_i^{(g)}.
	%	\end{aligned}
%\end{equation}
\begin{equation}
	\begin{aligned}
		H_{1}^{(g)}(-i\partial_{R_x}, \tilde{\bm{k}}) &= 2t_1\sin{(-i\partial_{R_x})} \gamma_1^{(g)}  + \left[m(R_x) - 2\tilde{t}_1 \cos{(-i\partial_{R_x})} -\sum_{i=2}^{d} 2\tilde{t}_i \cos{( \tilde{\bm{k}})} \right] \gamma_{d+1}^{(g)}  \\
		H_{d-1}^{(g)}( \tilde{\bm{k}}) &= \sum_{i=2}^{d}  2t_i \sin{( \tilde{\bm{k}})} \gamma_i^{(g)}.
	\end{aligned}
\end{equation}

By solving the real-space quasi-1d eigen equation:
\begin{equation}
	H_{1}^{(g)}(-i\partial_{R_x}) \ \mathcal{U}(R_x, \tilde{\bm{k}})
	= 0,
	\label{Eq_quasi_1d_eigen_equation}
\end{equation}
\begin{comment}
By solving the effective Dirac model along x-direction, e.g.,:
\begin{equation}
    (-i v_x \partial_x \sigma_x + m(x) \sigma_z) \ \mathcal{U}(R_x) = E^{\text{B}} \  {\mathcal{U}(R_x),
    \end{equation}
\end{comment}
we are able to obtain the TBMs in the momentum channel $\tilde{\bm{k}}$, with energy $E_{\tilde{\bm{k}}} =  \sum_{i=2}^{d}  2t_i \sin{( \tilde{\bm{k}})}$ and wave functions $\Psi^{\text{B}}(x)$. To solve the infinite-order differential equation Eq.~(\ref{Eq_quasi_1d_eigen_equation}), we plug in a trial function $\mathcal{U}_{r}(R_x, \tilde{\bm{k}}) = A_r e^{-R_x/\xi_{c,r}} u$ for region $R_x\ge 0$, where $u$ is a $N_f\times 1$ vector and $\xi_{c,r}$ denotes the localization length on the right region. It yields:
\begin{subequations} \label{Eq_quasi_1d_sol}
    \begin{equation}
        2\eta t_1 \sinh{\left(\xi_{c,r}^{-1}\right)} 
        + \left[m_r - 2\tilde{t}_1 \cosh{\left(\xi_{c,r}^{-1}\right)} 
        -\sum_{i=2}^{d} 2\tilde{t}_i \cos{\left(\tilde{\bm{k}}\right)} \right] = 0,
        \label{Eq_quasi_1d_sol_a}
    \end{equation}
    \begin{equation}
        i\gamma_1^{(g)}\gamma_{d+1}^{(g)} {u} = \eta {u}.
        \label{Eq_quasi_1d_sol_b}
    \end{equation}
\end{subequations}

\begin{comment}
    \textcolor{red}{For $d>2$?! $[\gamma_0\gamma_d, \gamma_{i\neq d}] = 0$}
\end{comment}
	
The solution to Eq.~(\ref{Eq_quasi_1d_sol_a}) yields the conventional length $\xi_{c,r}$:
\begin{equation}
    \xi_{c,r}^{-1} = 
    \begin{cases}
        \ln{\left(\frac{ -M  \pm \sqrt{M^2+4(t_1-\tilde{t}_1)(t_1+\tilde{t}_1)}}{2(t_1-\tilde{t}_1)}\right) }
        \quad & \text{for $\eta = 1$}, \\
        \ln{\left( \frac{M \pm \sqrt{M^2+4(t_1-\tilde{t}_1)(t_1+\tilde{t}_1)}}{2(t_1+\tilde{t}_1)}\right) } & \text{for $\eta = -1$},
    \end{cases}
    \label{Eq_xi_c}
\end{equation}
where $M = m_r-\sum_{i=2}^{d} 2\tilde{t}_i \cos{\left(\tilde{\bm{k}}\right)}$. The conventional length $\xi_{c,r}$ is generally complex with the real part $\Re(\xi_{c,r})>0$, representing an oscillatory decaying profile. For example, when $M=0$, the conventional length would be $\xi_{c,r} = 2/ \ln{\frac{t_1-\tilde{t}_1}{t_1+\tilde{t}_1}}$ or $\tilde{\xi}_{c,r} = 2/( \ln{\frac{t_1-\tilde{t}_1}{t_1+\tilde{t}_1}} + i\pi )$. The general solution in the right region is a linear combination of the two solutions:
\begin{equation}
    \mathcal{U}_{r}(R_x, \tilde{\bm{k}}) = \left(A_r e^{-R_x/\xi_{c,r}} + \tilde{A}_r e^{-R_x/\tilde{\xi}_{c,r}} \right) u,
    \label{Eq_envelop_vector_SM}
\end{equation}
which means that the conventional behavior of the TBMs is in general not an exponentially oscillatory decay. It will decay exponentially when only one solution to $\xi_{c,r}$ survives. For example, when $t_1 = \tilde{t}_1$, the conventional length can only be $\xi_{c,r} = 1/\ln{\frac{M}{2t_1}}$, resulting in a single exponentially decaying mode.

Similarly, for the left region $R_x<0$, the solution to Eq.~(\ref{Eq_quasi_1d_eigen_equation}) would be $\mathcal{U}_{l}(R_x, \tilde{\bm{k}}) = A e^{R_x/\xi_{c,l}} u$ where $\xi_{c,l}$ is also determined by Eq.~(\ref{Eq_xi_c}) with $m_r$ replaced by $m_l$. Then combining the solutions in the two regions, we write down the complete wave functions for the TBMs
\begin{equation}
    \begin{aligned}
        \Psi^{\text{B}}(x) = \sum^{N_f}_{\alpha} \sum_{R_x} u_{\alpha} \left[ \left(A_r e^{-R_x/\xi_{c,r}} + \tilde{A}_r e^{-R_x/\tilde{\xi}_{c,r}} \right) \Theta(R_x) + \left(A_l e^{-R_x/\xi_{c,l}} + \tilde{A}_l e^{-R_x/\tilde{\xi}_{c,l}} \right) \Theta(-R_x-1)\right] \mathcal{W}_{\alpha}(x-R_x,\tilde{\bm{k}})
    \end{aligned}
    \label{Eq_edge_states_wave_functions4},
\end{equation}
%
%\begin{equation}
%	\begin{aligned}
    %		\Psi^{\text{B}}(x) = \sum^{N_f}_{\alpha} \sum_{R_x=0} {u}_{\alpha} \left( A_r e^{-R_x/\xi_{c,r}}\Theta(R_x) + A_l e^{R_x/\xi_{c,l}} \Theta(-R_x-1)\right) \mathcal{W}_{\alpha}(x-R_x,\tilde{\bm{k}})
    %	\end{aligned}
%	\label{Eq_edge_states_wave_functions4},
%\end{equation}
with ${u}_{\alpha}$ being the eigenvector in Eq.~(\ref{Eq_quasi_1d_sol_b}). When it is applied with open boundary conditions in a certain direction, the TBMs would take a similar form to Eq.~(\ref{Eq_edge_states_wave_functions4}), except that the two terms in the square bracket will be arranged at opposite boundaries of the system and form two boundary modes individually. In this case, the object we care about would no longer be the single OBC wave functions, but rather a rearranged form of the wave functions located at opposite boundaries, because they essentially share the entire Wannier orbitals. Specifically, let us denote the left and right boundary modes as $\psi_L(x)$ and $\psi_R(x-x_R)$ with $\psi_L(0)/\psi_R(0) \equiv q$. Then we rearrange the wave functions as $\tilde{\Psi}^{\text{B}}(x) = \sqrt{\frac{1}{1+\mathcal{N}_q}} (\psi_L(x)+q\psi_R(x)\Theta(-x-1))$, where $\mathcal{N}_q \equiv \sum_{x=-\infty}^{-1} |q\psi_R(x)|^2$ to normalize $\tilde{\Psi}^{\text{B}}(x)$. For example, in the case with an asymmetric Wannier orbital, one has $\psi_R(0) = 1, \tilde{\psi}^{\text{B}}(x) = \psi_L(x)$. As a result, the remainder of the discussions follows with $\Psi^{\text{B}}(x)$ in Eq.~(\ref{Eq_edge_states_wave_functions4}) or the rearranged wave functions $\tilde{\Psi}^{\text{B}}(x)$.

%%%%%%%%%%%%%%%%%%%%%%%%%%%%%%%%%%%%%%%%%%%%%%%

\subsection{Two phases of behaviors of TBMs}

\begin{comment}
The envelop vector $\mathcal{U}_{\alpha}(R_x, \tilde{\bm{k}})$ satisfies Eq.\eqref{}
The exact solutions in general comprise 2 exponentially decaying modes, i.e.,
\begin{equation}
   \begin{aligned}
        \mathcal{U}_{\alpha}(R_x, \tilde{\bm{k}}) = {u}_{\alpha} \left( e^{-R_x/\xi_c} + e^{-R_x/\tilde{\xi}_c}\right)
    \end{aligned}
    \label{Eq_envelop_func_structure}.
\end{equation}
Here, the conventional lengths $\xi_c$ and $\tilde{\xi}_c$, are determined by the low-energy effective Hamiltonian, and are independent of quantum geometry. $\xi_c$ in general adopts a complex form with a positive real part, manifesting an oscillatory decaying behavior, as evident in Fig.~\ref{Fig_Lieb_QWZ_OBC_spectrum_wavefunc_QSQM}(g) where the conventional oscillatory behavior is denoted in dashed lines. More discussions on Eq.~(\ref{Eq_envelop_func_structure}) is presented in Supplemental Material \cite{}. For simplicity, in the following we consider the case $t_i=\tilde{t}_i$ in which the two exponential branches are identical with $\xi_c = \tilde{\xi}_c$.
\end{comment}

Wannier orbitals are not unique, depending on the specific gauge imposed for the Bloch vectors. However, the Wannier orbitals serving as the bases for a physical observable state cannot be arbitrary. Consider the limit $\xi_c \rightarrow 0$, the TBMs wave functions in Eq.~(\ref{Eq_edge_states_wave_functions4}) becomes a simple summation of the Wannier orbitals at the boundary site:
\begin{equation}
    \begin{aligned}
        \Psi^{\text{B}}(x) \overset{\xi_c \rightarrow 0}{=}  \sum_{\alpha} \sum_{R_x} {u}_{\alpha} \delta_{R_x, 0} \mathcal{W}_{\alpha}(x-R_x,\tilde{\bm{k}})
        = \sum_{\alpha} \sum_{R_x} {u}_{\alpha} \mathcal{W}_{\alpha}(x-0,\tilde{\bm{k}}).
    \end{aligned}
\end{equation}
As eigen-states, the Wannier orbitals without topological obstruction naturally adopt an exponentially decaying form \cite{PhysRev.115.809,PhysRevLett.86.5341}.
%up to a polynomial factor.
% \textcolor{red}{Discuss Wannier Obstruction?}.
Therefore, we consider the exponentially localized Wannier orbitals, which are typically of the following form:
\begin{equation}
    \begin{aligned}
        \mathcal{W}_{\alpha}(x-R_x,\tilde{\bm{k}}) &= A_{\alpha} \delta(x-R_x) \\ 
        &+ B_{\alpha} e^{-(x-R_x)/{\xi_g}} \Theta\left(x-(R_x+1)\right),
    \end{aligned}
    \label{Eq_expo_localized_wannier_func}
\end{equation}
where $\Theta(x)$ is the Heaviside step function and $\xi_g$ is the decay length of the exponentially localized Wannier orbitals. Without loss of generality, we have assumed an asymmetric form of Wannier orbitals that decay rightward. $\xi_g$, $A_{\alpha}$ and $B_{\alpha}$ are also implicitly dependent on $\tilde{\bm{k}}$. In general, the coefficients satisfy $|A_{\alpha}| > |B_{\alpha}|$. When we plug Eqs.(\ref{Eq_envelop_vector_SM}) (we consider a simplified case with $\xi_c = \tilde{\xi}_c$)
%For simplicity, we consider the case $t_i=\tilde{t}_i$ in which the two exponential branches are identical
and (\ref{Eq_expo_localized_wannier_func}) into Eq.~(\ref{Eq_edge_state_wavefunc_appendix}), the boundary-mode wave functions become ($\xi_c \neq \xi_g$):
\begin{equation}
    \begin{aligned}
        \Psi^{\text{B}}(x,\tilde{\bm{k}}) 
        = A_c e^{-x/{\xi_c}} + A_{g} e^{-x/{\xi_g}},
    \end{aligned}
    \label{Eq_edge_states_wave_functions_method}
\end{equation}
with
\begin{subequations}
    \begin{eqnarray}
        A_c &=& \sum_{\alpha} {u}_{\alpha} \left[ A_{\alpha} - B_{\alpha} (1-e^{-(\frac{1}{\xi_c}-\frac{1}{\xi_g})})^{-1} \right] \label{Eq_A_c__A_QM_a}  \\
        A_{g} &=& \sum_{\alpha} {u}_{\alpha} B_{\alpha} \left(1-e^{-(\frac{1}{\xi_c}-\frac{1}{\xi_g})}\right)^{-1}. \label{Eq_A_c__A_QM_b}
    \end{eqnarray}
    \label{Eq_A_c__A_QM}
\end{subequations}
%\begin{equation}
%	\begin{aligned}
    %		&A_c = \sum_{\alpha} {u}_{\alpha} \left(A_{\alpha} - B_{\alpha} \left(1-e^{-(\frac{1}{\xi_c}-\frac{1}{\xi_g})}\right)^{-1} \right) \\ 
    %		&A_{g} = \sum_{\alpha} {u}_{\alpha} B_{\alpha} \left(1-e^{-(\frac{1}{\xi_c}-\frac{1}{\xi_g})}\right)^{-1}.
    %	\end{aligned}
%	\label{Eq_A_c__A_QM}
%\end{equation}
Near the boundary, the states initially undergo a conventional oscillatory decay with diminishing oscillation amplitude. Beyond this region, once the oscillation fades, a pure exponential decay emerges.
However, these two behaviors may not always be observed simultaneously, as their prominence depends on the relative scales of the conventional length $\xi_c$ and the geometric length $\xi_g$. When the conventional length dominates, i.e., $\xi_c \gg \xi_g > 0$, Eq.~(\ref{Eq_A_c__A_QM}) simplifies to:
\begin{equation}
    A_c = \sum_{\alpha} {u}_{\alpha} \left(A_{\alpha} + B_{\alpha}\xi_g \right), A_{g} = - \sum_{\alpha} {u}_{\alpha} B_{\alpha}\xi_g.
\end{equation}
In this case, the first term in Eq.~(\ref{Eq_edge_states_wave_functions_method}) overwrites the second, effectively masking the geometric behavior.
Conversely, the geometric length will be manifested when $\xi_g > \xi_c$ and it becomes the dominant length scale when $\xi_g \gg \xi_c > 0$. In this regime, Eq.~(\ref{Eq_A_c__A_QM}) becomes:
\begin{equation}
    A_c = \sum_{\alpha} {u}_{\alpha} \left(A_{\alpha} - B_{\alpha}\xi_c \right), A_{g} = \sum_{\alpha} {u}_{\alpha} B_{\alpha}\xi_c.
\end{equation}
Examining the two terms in Eq.~(\ref{Eq_edge_states_wave_functions_method}), it is evident that, at long range, the boundary modes are ultimately governed by the geometric behavior. However, over the first few sites near the boundary, either term may prevail. If the conventional decay initially dominates, it will soon transition to a slower exponential decay dictated by $\xi_g$. When $\xi_c \rightarrow 0$, the topological boundary modes only follow the geometric behavior. 

%%%%%%%%%%%%%%%%%%%%%%%%%%%%%%%%%%%%%%%%%%555
\subsection{Universal bound to the spatial extent of TBMs}
In the flat band limit with $\xi_c \rightarrow 0$, we look at the spread function of the TBMs:
\begin{equation}
    \begin{aligned}
        \Omega^{x}_{\Psi^{\text{B}}} 
        \equiv 
        &\int_{x} (\left(\Delta{\hat{x}} \right)_{|\Psi^{\text{B}}(x)|^2}^2 \diff x  \\
        = &\int_{x} {\Psi^{\text{B}}}^{\dagger}(x) \hat{x}^2 \Psi^{\text{B}}(x) \diff x 
        - 	\left|\int_{x} {\Psi^{\text{B}}}^{\dagger}(x) \hat{x} \Psi^{\text{B}}(x) \diff x \right|^2 \\
        \ge & \sum_{\alpha, \alpha'} {u}^*_{\alpha} {u}_{\alpha'} \int_{x} \mathcal{W}_{\alpha}^{\dagger}(x-0,\tilde{\bm{k}}) \hat{x}^2 
        \mathcal{W}_{\alpha'}(x-0,\tilde{\bm{k}}) \diff x  -	 \sum_{R_x} \left| \sum_{\alpha, \alpha'} {u}^*_{\alpha} {u}_{\alpha'}   \int_{x} \mathcal{W}_{\alpha}^{\dagger}(x-R_x,\tilde{\bm{k}}) \hat{x} 
        \mathcal{W}_{\alpha'}(x-0,\tilde{\bm{k}}) \diff x \right|^2 
        \\
        = & \sum_{\alpha, \alpha'} {u}^*_{\alpha} {u}_{\alpha'} \frac{1}{N_x} \sum_{k_x} \partial_{k_x} \mathcal{F}_{\bm{k},\alpha}^{\dagger} \partial_{k_x} \mathcal{F}_{\bm{k},\alpha}  
        - \sum_{\alpha, \alpha'} \sum_{\beta, \beta'} {u}^*_{\alpha} {u}_{\alpha'} {u}_{\beta} {u}^*_{\beta'}
        \frac{1}{N_x} \sum_{k_x} \partial_{k_x} \mathcal{F}_{\bm{k},\beta'}^{\dagger}  \mathcal{F}_{\bm{k},\beta}  \mathcal{F}_{\bm{k},\alpha}^{\dagger}  \partial_{k_x} \mathcal{F}_{\bm{k},\alpha'}
        \\
        = & \frac{1}{N_x}  \sum_{\alpha, \alpha'} \sum_{k_x} \partial_{k_x} \mathcal{F}^{\dagger}_{\bm{k}, \alpha} \left({u}^*_\alpha {u}_{\alpha'} - \sum_{\beta, \beta'} {u}^*_{\beta'} {u}_{\alpha'} {u}_{\beta} {u}^*_{\alpha}
        \mathcal{F}_{\bm{k}, \beta} \mathcal{F}^{\dagger}_{\bm{k}, \beta'} \right) \partial_{k_x} \mathcal{F}_{\bm{k}, \alpha'}   \\
        = & \frac{1}{N_x}  \sum_{\alpha, \alpha'} {u}^*_\alpha {u}_{\alpha'} \sum_{k_x} \partial_{k_x} \mathcal{F}^{\dagger}_{\bm{k}, \alpha} 
        \left(1 - \sum_{\beta, \beta'} {u}^*_{\beta'}  {u}_{\beta} 
        \mathcal{F}_{\bm{k}, \beta} \mathcal{F}^{\dagger}_{\bm{k}, \beta'} \right) \partial_{k_x} \mathcal{F}_{\bm{k}, \alpha'}   \\
        \equiv &  \frac{1}{N_x}  \sum_{k_x} \mathcal{G}^{0}_{xx}(\bm{k}), \\
    \end{aligned}
    \label{Eq_Edgestates_localization_function}
\end{equation}

where in the fourth step we have used:
\begin{equation}
    \begin{aligned}
        &\int_{x} \mathcal{W}_{\alpha}^{\dagger}(x-R_x',\tilde{\bm{k}}) \hat{x} 
        \mathcal{W}_{\beta}(x-R_x,\tilde{\bm{k}}) \diff x \\
        = & \frac{1}{N_x^2} \int_x  \sum_{k_x, k_x'} e^{-ik_x (x-R_x')} e^{ik_x' (x-R_x)} x \mathcal{F}_{\bm{k},\alpha}^{\dagger} \mathcal{F}_{\bm{k}',\beta} \diff x\\
        = & \frac{1}{N_x^2} \int_x  \sum_{k_x, k_x'} (i \partial_{k_x} e^{-i(k_x-k_x')x}) e^{ik_x R_x'} e^{-ik_x' R_x} \mathcal{F}_{\bm{k},\alpha}^{\dagger} \mathcal{F}_{\bm{k}',\beta} \diff x \\
        = &  -\frac{i}{N_x^2}   \sum_{k_x, k_x'}  \int_x  e^{-i(k_x-k_x')x} \diff x \ e^{ik_xR_x'} e^{-ik_x' R_x} (\partial_{k_x} \mathcal{F}_{\bm{k},\alpha}^{\dagger}) \mathcal{F}_{\bm{k}',\beta} \\
        = & -\frac{i}{N_x} \sum_{k_x}  e^{ik_x (R_x'-R_x)}  (\partial_{k_x} \mathcal{F}_{\bm{k},\alpha}^{\dagger})  \mathcal{F}_{\bm{k},\beta}, \\
        = & \frac{i}{N_x} \sum_{k_x}  e^{ik_x (R_x'-R_x)}  \mathcal{F}_{\bm{k},\alpha}^{\dagger} \partial_{k_x} \mathcal{F}_{\bm{k},\beta} \\
    \end{aligned}
    \label{Eq_wannier_relation1}
\end{equation}
and
\begin{equation}
    \begin{aligned}
        \int_{x} \mathcal{W}_{\alpha}^{\dagger}(x-R_x',\tilde{\bm{k}}) \hat{x} ^2
        \mathcal{W}_{\beta}(x-R_x,\tilde{\bm{k}}) \diff x 
        =  -\frac{1}{N_x} \sum_{k_x}  e^{ik_x (R_x'-R_x)}  \mathcal{F}_{\bm{k},\alpha}^{\dagger} \partial_{k_x}^2 \mathcal{F}_{\bm{k},\beta} \\
    \end{aligned}.
    \label{Eq_wannier_relation2}
\end{equation}

After the inequality scaling in Eq.~(\ref{Eq_Edgestates_localization_function}), we obtain a simple form of the lower bound for the spread of the boundary modes. It is the xx-component of the BZ-integrated quantum metric tensor $\mathcal{G}^{0}_{ab}(\bm{k})$, which is defined as:
\begin{equation}
    \begin{aligned}
        \mathcal{G}^{0}_{ab}(\bm{k}) 
        &\equiv \frac{1}{2} \left[ \sum_{\alpha, \alpha'}^{N_f} {u}^*_\alpha {u}_{\alpha'} \partial_{k_a} \mathcal{F}^{\dagger}_{\bm{k}, \alpha} \left(1 - \sum_{\beta, \beta'}^{N_f} {u}^*_{\beta'}  {u}_{\beta} 
        \mathcal{F}_{\bm{k}, \beta} \mathcal{F}^{\dagger}_{\bm{k}, \beta'} \right) \partial_{k_b} \mathcal{F}_{\bm{k}, \alpha'} + (a \leftrightarrow b) \right] \\
        &=  \Re \left[ \partial_{k_a} \Psi_0(\bm{k}) \left(1 - \Psi_0(\bm{k}) \Psi_0^{\dagger}(\bm{k}) \right) \partial_{k_b} \Psi_0(\bm{k}) \right],
    \end{aligned}
    \label{Eq_SpreadLowerBound_QM}
\end{equation}
where we have defined flat-band eigen-states $\Psi_0(\bm{k}) \equiv \sum_{\alpha}^{N_f} u_{\alpha} \mathcal{F}_{\bm{k}, \alpha}$. From Eq.~(\ref{Eq_distance_of2_vectors}) one can readily tell that $\mathcal{G}^{0}_{ab}(\bm{k})$ is the metric that measures the quantum distance between two infinitesimally adjacent states $\Psi_0(\bm{k})$ and $\Psi_0(\bm{k} + \diff \bm{k})$. As $\Psi_0(\bm{k})$ is a vector in the degenerate flat-band manifold $S_{\mathcal{F}}$, its quantum metric can be related to the non-Abelian quantum metric of the degenerate manifold, as given in Eq.~\eqref{Eq_Relation_Abelian_and_nonAbelian_QM}. With that, we find a further inequality between the xx-component of the two metric:
\begin{equation}
    \begin{aligned}
        \mathcal{G}^{0}_{xx}(\bm{k}) 
        &\equiv \sum^{N_f}_{\alpha \alpha'} {u}_{\alpha}^{\dagger} \mathcal{G}^{\alpha\alpha'}_{xx}(\bm{k}) {u}_{\alpha'} + \left| \partial_x \Psi_{0_{\perp},\bm{k}}^{\dagger} \Psi_{0,\bm{k}}  \right|^2 \\
        &\ge \sum^{N_f}_{\alpha \alpha'} {u}_{\alpha}^{\dagger} \mathcal{G}^{\alpha\alpha'}_{xx}(\bm{k}) {u}_{\alpha'},
    \end{aligned}
    \label{Eq_inequality_G0_ge_non_Abelian_QM}
\end{equation}
which, together with the inequality in Eq.~(\ref{Eq_Edgestates_localization_function}), amounts to the core result of this work:
\begin{equation}
    \Omega^{x}_{\Psi_{\alpha}^{\text{B}}} \ge  \frac{a}{2\pi} \sum^{N_f}_{\alpha \alpha'} \int_{\text{BZ}}  {u}_{\alpha}^{\dagger} \mathcal{G}^{\alpha\alpha'}_{xx}(\bm{k}) {u}_{\alpha'} \diff k_x,
    \label{Eq_inequality_LocalizationFunc_ge_nonAbelian_QM}
\end{equation}
where ${u}_{\alpha}$ is an eigenvector of $\gamma^{g}_1 \gamma^{g}_{d+1}$, and $\mathcal{G}^{\alpha\alpha'}_{xx}(\bm{k})$ is the non-Abelian quantum metric of the original $N_f$ degenerate flat bands. 

More generally, by defining the $i$-directional quantum metric length (QML) as:
\begin{equation}
        \xi_{QM,i, \tilde{\bm{k}}} \equiv  \sum^{N_f}_{\alpha \alpha'} \frac{1}{2\pi}  \int_{\text{BZ}}  {u}_{\alpha}^{\dagger} \mathcal{G}^{\alpha\alpha'}_{ii}(\bm{k}) {u}_{\alpha'} \diff k_i,
    \label{Eq_QML_definition_SM}
\end{equation}
the $i$-directional spread function $\Omega^{i}_{\Psi_{\alpha}^{\text{B}}}$ will be lower bounded by the corresponding QML (up to a lattice constant) as: $\Omega^{i}_{\Psi_{\alpha}^{\text{B}}} \ge a\xi_{QM,i, \tilde{\bm{k}}} $, with ${u}_{\alpha}$ being an eigenvector of $\gamma^{g}_i \gamma^{g}_{d+1}$.

%%%%%%%%%%%%%%%%%%%%%%%%%%%%%%%%%
%\subsection{Discussions on the definition of the QML}
To better understand above definition for QML, we now add a few remarks. First, as a concrete example, let us consider 2D TIs with coupling matrix $H_{2}^{(1)}$ under the \(y\)-OBC. In this case, ${u}_{\alpha}$, the eigenvector of $i\sigma_3^{(2)}\sigma_2^{(2)} \propto \sigma_1$, would have uniform components ${u}_{\alpha} \propto (1,1)/\sqrt{2}$, resulting in a rather simplified expression for the QML: $\xi_{QM,x,k_y} = \frac{1}{2\pi N_f}   \sum^{N_f}_{\alpha \alpha'}  \int_{\text{BZ}} \mathcal{G}^{\alpha\alpha'}_{xx}(\bm{k}) \diff k_x$. Additionally, for topological superconductors with a single flat band in the normal state, such as the Lieb-$p+ip$ model provided in Supplementary Note III, the doubly degenerate flat bands arise from enforced particle-hole symmetry. As a result, the integrand in Eq.~(\ref{Eq_QML_definition_SM}) corresponds directly to the Abelian quantum metric of the single normal flat band, $\mathcal{G}^f_{ab}$, as $\frac{1}{N_f} \sum_{\alpha\alpha'} {u}_{\alpha}^{\dagger} \mathcal{G}^{\alpha\alpha'}_{ab} {u}_{\alpha'} = \mathcal{G}^f_{ab}$. This ensures that Eq.~(\ref{Eq_QML_definition_SM}) aligns with the QML definition in \cite{guo2024majoranazeromodesliebkitaev}.

%%%%%%%%%%%%%%%%%%%%%%%%%%%%%%%%%%%%%%%%%%%%%%%
\section{Supplementary Note III: Constructed topological flat-band models}
\begin{figure}[htbp]
    \centering
    \includegraphics[width=0.5\linewidth]{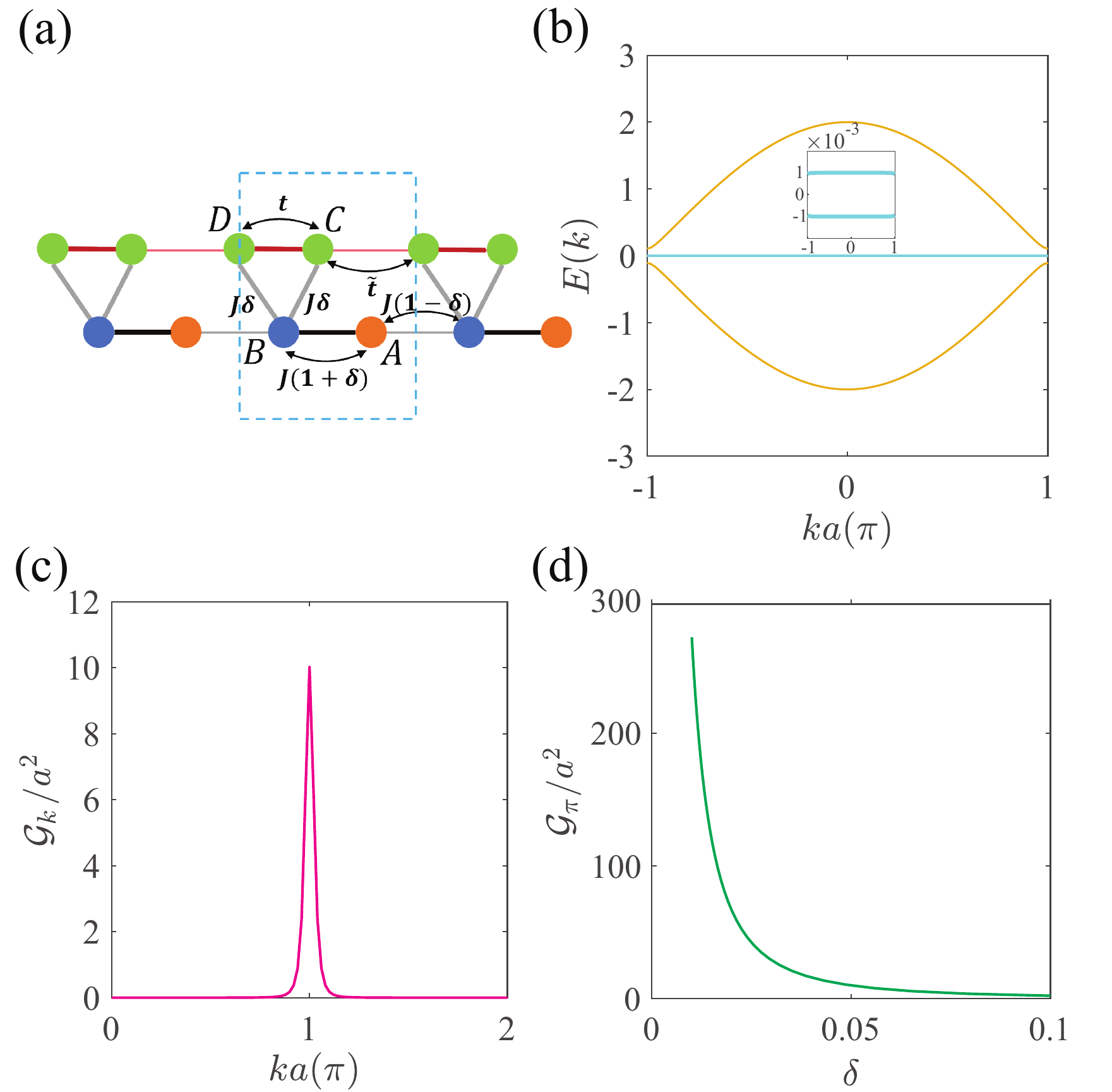}
    \caption{(a) The real-space lattice structure of the Lieb-SSH model. The green sites represent the C,D orbital forming a SSH chain. (b) The band structure of the Lieb-SSH model featuring two dispersive bands (yellow) and two central flat bands (cyan), with the inset providing a zoomed-in view of the flat bands. The parameters are $J=1$, $\tilde{t}= 2\times10^{-4}J$, $t = 0$, 
    %	$t = \tilde{t}$ 
    and $\delta = 0.05$. (c) The BZ distribution of the quantum metric tensor $\mathcal{G}_k$ for a single flat band. (d) The quantum metric at $ka = \pi$ is tunable by the parameter $\delta$.}
    \label{Fig_Lieb_SSH_Band_QM}
\end{figure}

\subsection{\label{sec_3-2-2} Lieb-SSH model}
With the Lieb-QWZ model introduced in the main text, we consider the channel $k_2 = 0$, and obtain an effective 1d model.
%With $d = 1$, $g = 1$, expand the Eq.~(\ref{Eq_Dirac_hamilt_H_dg_continuum}) around $k = \pi/2$ on square lattice, resulting in:
%\begin{equation}
%	H_1^{(1)}(k) =  2t\sin(k_1a)\sigma_x + (m - 2t\cos(k_1a)) \sigma_y,
%	\label{Eq_SSH_hamilt}
%\end{equation}
Take $d = 1$ and $g = 1$, we have the coupling matrix as:
\begin{equation}
    H_1^{(1)}(k) = [t + \tilde{t} \cos(ka)] \sigma_x  +  \tilde{t} \sin(ka)\sigma_y,
    \label{Eq_SSH_hamilt}
\end{equation}
which itself is the SSH model. Embedding Eq.~(\ref{Eq_SSH_hamilt}) into the variant Lieb model yields a 1D Lieb-SSH model. When $S_{2, \bm{k}} = \left[(1+\delta) - (1-\delta)e^{ik_2a}, 0\right]$, the Lieb-SSH model belongs to the AIII class in the AZ classification. It has chiral symmetry $\hat{C} H_{AIII} \hat{C}^{-1} = -H_{AIII}$ with the chiral operator represented as
%\begin{equation}
%	\begin{split}  
    %		D(\hat{C}) =  \begin{pmatrix}
        %			1 & 0 & 0 & 0 \\ 
        %			0 & -1& 0 & 0 \\ 
        %			0 & 0 & 0 & 1 \\ 
        %			0 & 0 & 1 & 0 \\ 
        %		\end{pmatrix}.
    %	\end{split} 
%\end{equation}
$D(\hat{C}) = \operatorname{diag}(1, -1, -1, 1)$.
A lattice structure for the Lieb-SSH model is shown in Fig.~\ref{Fig_Lieb_SSH_Band_QM}(a). Similar to the Lieb-QWZ model, the Lieb-SSH model also contains four bands with two flat bands in the middle, as shown in Fig.~\ref{Fig_Lieb_SSH_Band_QM}(b), where the flat bands are zoomed in the inset. In Fig.~\ref{Fig_Lieb_SSH_Band_QM}(c), we calculate and plot the Abelian quantum metric of one flat band, which peaks at the point $k=\pi$. By tuning $\delta$, the quantum metric at $k=\pi$ varies accordingly, as presented in Fig.~\ref{Fig_Lieb_SSH_Band_QM}(d), as expected.

%%%%%%%%%%%%%%%%%%%%%%%%%%%%%%%%%%%%%%%%%%%%%%%%%

\subsection{\label{sec_3-2-3} Lieb-$p+ip$ model}

%%%%%%%%%%%%%%%%%%%%%%%%%%%%%
\begin{figure}[hbp]
    \centering
    \includegraphics[width=0.5\linewidth]{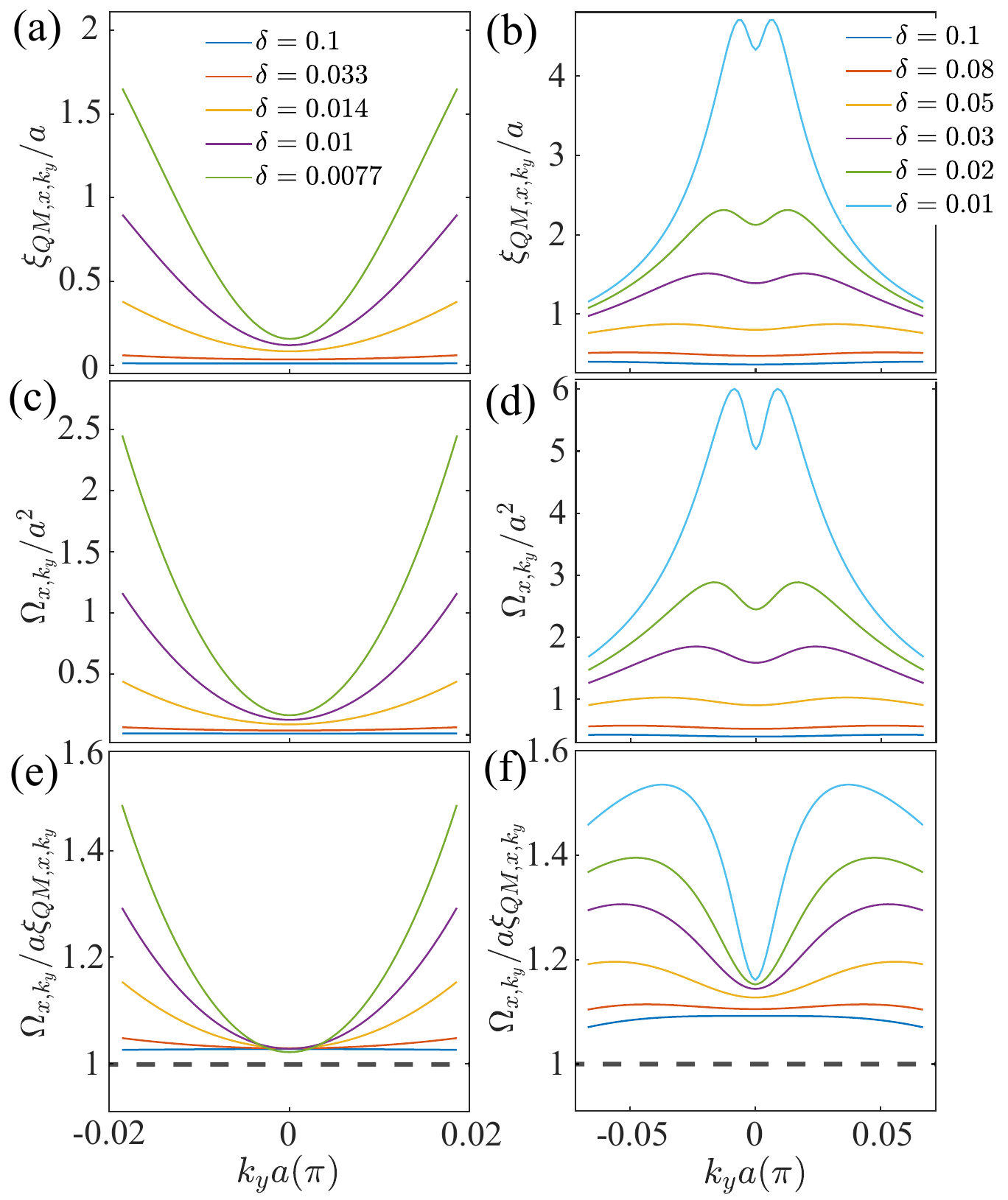}
    \caption{(a, b) The \(x\)-direction QML $\xi_{QM, x, k_y}$ versus different $k_y$ channels for (a) the Lieb-QWZ model and (b) the Lieb-$p+ip$ model, tunable via the parameter $\delta$, with shared parameters $J=1$, $\tilde{t}= 2\times10^{-4}J$, $m = 2\tilde{t}$, $t = \Delta = \tilde{t}$, $\delta = 0.05$, and $\alpha = 0.1$ for the Lieb-QWZ model. 
    %(Here we choose $\alpha$ to be consistent with previous plots; however, a larger $\alpha$ can increase the QML by several orders of magnitude, while simultaneously altering the band structure of the flat bands in Fig.~\ref{Fig_Lieb_QWZ_OBC_spectrum_wavefunc_QSQM}(b) and (f)). 
    (c, d) The \(x\)-directional spread function $\Omega_{x,k_y}$ for (c) the Lieb-QWZ model and (d) the Lieb-$p+ip$ model; The ratio $\Omega_{x,k_y}/a\xi_{QM}$ where $a$ is the lattice constant, is consistently bounded below by 1 (black dashed line) in both (c) the Lieb-QWZ model and (d) the Lieb-$p+ip$ model, in excellent agreement with Eq.~\eqref{Eq_inequality_LocalizationFunc_ge_nonAbelian_QM}.}
    \label{Fig_Lieb_QWZ_QSQM}
\end{figure}

When the replicated sublattices are introduced by particle-hole symmetry. We can further realize a 2D topological superconductor (TSC). Specifically, a 2D Lieb-$p+ip$ model is realized when imposing:
\begin{equation}
    \begin{aligned}
        S_{1,\bm{k}} &= \left[ (1+\delta)+(1-\delta)e^{ik_1a}\right]\sigma_z \\
        S_{2, \bm{k}} &= \left[(1+\delta) - (1-\delta)e^{ik_2a}\right] \mathcal{I}_{1\times 2^g} \sigma_z \\
        H_2^{(1)}(\bm{k}) &=  2\Delta \sin(k_1a)\sigma_x + 2\Delta\sin(k_2a)\sigma_y + M(\bm{k}) \sigma_z,
    \end{aligned}
    \label{Eq_S1_S2_Lieb_TSC} 
\end{equation}
where $\sigma_{x,y,z}$ are Pauli matrices in the particle-hole basis, and $M(\bm{k}) = m - 2t(\cos(k_1a) + \cos(k_2a))$. Here, we also require $\operatorname{max}(t, \Delta, m) \ll  \delta J$ to meet the flatness condition. 
%And $\lambda_{\bm{k}=0} = \delta \ll 1$ satisfying Eq.~(\ref{Eq_recover_low_energy_model_condition}) so that the low-energy model is well-described by $H_2^{(1)}(\bm{k} \rightarrow 0)$. 
At the $M$-point, where the gap between the flat bands and higher dispersive bands is located, the local quantum geometry indicator $\lambda_{M} = 1$ with the band gap $\Delta_g \sim \delta J$. Therefore, the M-point non-Abelian quantum metric can be enhanced by reducing $\Delta_g$ via parameter $\delta$ according to our framework. A general BdG Hamiltonian should contain pairings among all orbitals, so a realistic coupling matrix for the topological superconductor takes $H_{\text{c}}(\bm{k}) = \mathcal{I}_3 \otimes H_2^{(1)}(\bm{k})$ with the identity matrix $\mathcal{I}_3$ in the orbital basis. With this enlarged degrees of freedom, we no longer invoke the parameter $\alpha$ in the Lieb-QWZ model, since the band structure of the flat bands in this model is well described by $H_2^{(1)}(\bm{k})$ throughout the BZ. 
%The model exhibits a similar band structure as the Lieb-QWZ model so is not displayed here. The quantum geometry property of the Lieb-$p+ip$ model will be inspected later.
The 2D Lieb-$p+ip$ features six Bogoliubov bands with two flat bands in the middle.

To calculate the QML of the flat bands in the Lieb-$p+ip$ model, we need to turn off the coupling term by setting $\Delta =t=m=0$. In this case, the flat bands become degenerate, enforced by particle-hole symmetry. For a better visual comparison, in Fig.~\ref{Fig_Lieb_QWZ_QSQM}(a) and (b), we present the QML for both the Lieb-QWZ model and the Lieb-$p+ip$ model.
The QML for both the Lieb-QWZ model and the Lieb-$p+ip$ model is tunable via the parameter $\delta$. To verify the relation in Eq.~(\ref{Eq_inequality_LocalizationFunc_ge_nonAbelian_QM}), we compute the \(x\)-OBC spread function $\Omega_{x, k_y}$ of the in-gap modes for both models near the $\Gamma$-channel ($\tilde{\bm{k}}=0$), where the conventional length is negligible. The results are presented in Fig.~\ref{Fig_Lieb_QWZ_QSQM}(c) and (d). Strikingly, both the Lieb-QWZ and the Lieb-$p+ip$ models exhibit a similar trend between the spread $\Omega_{x,k_y}$ and QML $\xi_{x,k_y}$ over the momentum channel $k_y$, revealing an unambiguous origin of the long-range behavior of the flat-band TBMs -- the exponentially localized Wannier orbitals. The ratio $\Omega_{x,k_y}/a\xi_{x,k_y}$, shown in Fig.~\ref{Fig_Lieb_QWZ_QSQM}(e) and (f), clearly demonstrates that the QML sets a lower bound on the spread of flat-band TBMs.

\subsection{\label{Sec_TBG} Ten-band model for magic-angle twisted bilayer graphene}
\begin{figure}[htbp]
    \centering
    \includegraphics[width=1\linewidth]{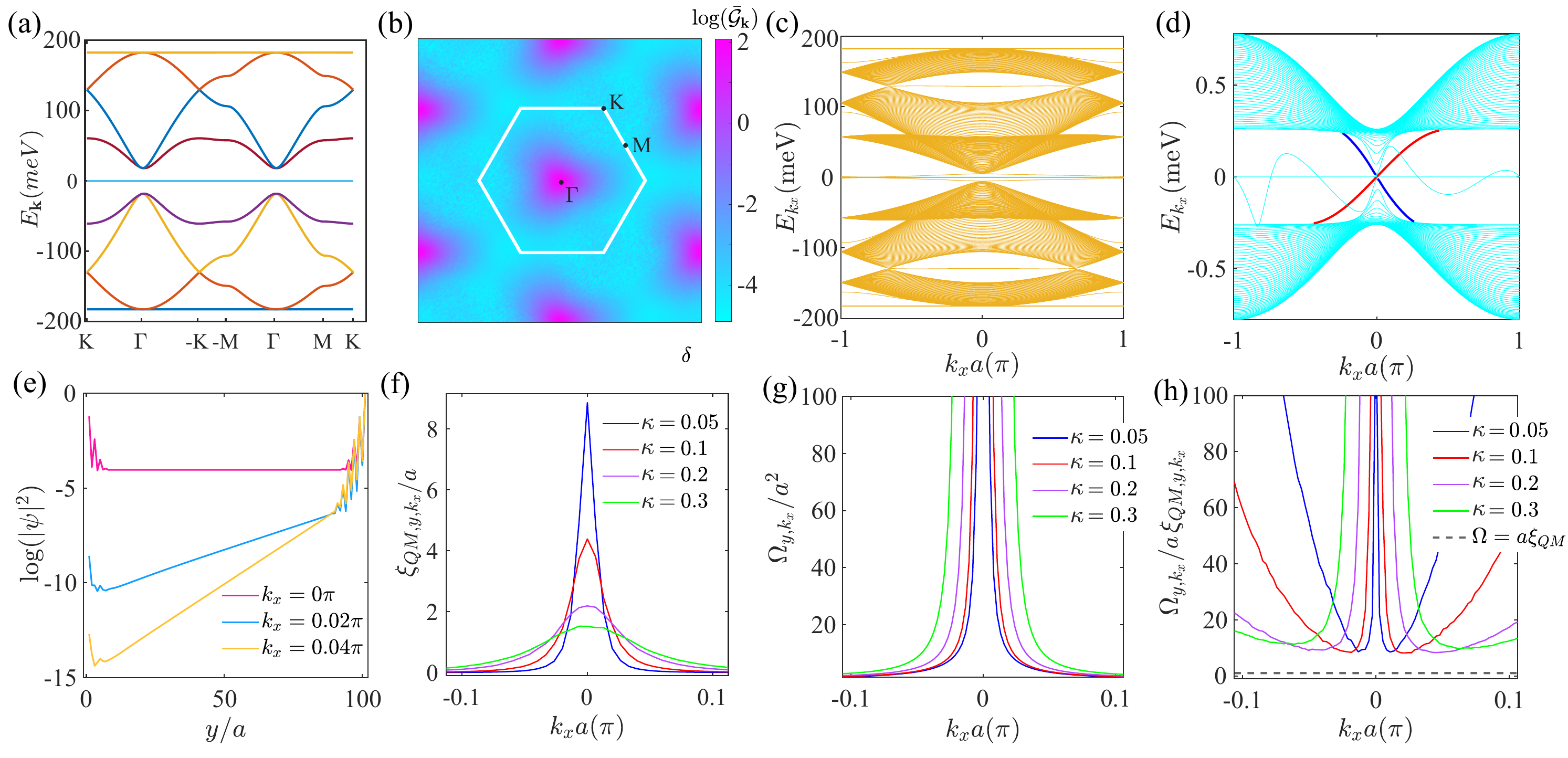}
    \caption{(a) The band structure for the 10-band model in the chiral limit, featuring two exact flat bands in the middle, with the parameters provided in \cite{PhysRevB.99.195455}. 
        (b) Logarithmic plot of the BZ distribution for the traced non-Abelian quantum metric tensor $\bar{\mathcal{G}}_{\bm{k}} \equiv \frac{1}{N_f}\sum_{\alpha\alpha'} \operatorname{Tr}[{\mathcal{G}_{ab}^{\alpha\alpha'}}(\bm{k})]$, revealing a hotspot region with threefold rotation symmetry. (c) Under $y$-OBC, the full OBC spectrum (orange) sandwiches two topological flat bands (cyan), zoomed in (d), with in-gap TBMs highlighted in red and blue. The flat band coupling parameters are $\tilde{t}= 1\times10^{-3}J=0.13$ meV, $m = 2\tilde{t}$, $t = \tilde{t}$ and $\kappa = 0.3$, and applied to subsequent panels unless otherwise specified. (e) The TBMs near the $\Gamma$-channel exhibit both the conventional oscillatory decay and an geometric behavior controlled by the QML. Parameter $t = 0.6\tilde{t}$. (f) The \(y\)-directional QML distribution around the $\Gamma$-channel, controlled by the gap-closing parameter $\kappa$. (g) The spread of the TBMs, also tunable by $\kappa$, aligns with the QML behavior away from the $\Gamma$-channel, where the conventional length is suppressed. (h) The ratio $\Omega_{y,k_x}/a\xi_{y, k_x}$ of the spread to the QML is well bounded below by 1.}
    \label{Fig_MATBG_band_QM_wavefunc_QS}
\end{figure}

Magic angle twisted bilayer graphene (MATBG) is a pioneering 2D platform renowned for its fascinating correlated phenomena, which is driven by the suppressed kinetic energy in the flat bands. This establishes MATBG a versatile arena for harnessing the rich flat-band physics \cite{cao2018unconventional}. One of the questions that has interested the community for a long time is the role of quantum geometry in shaping the MATBG physics. In this subsection, we dig further into the intricate interplay of topology and the quantum metric of the flat bands in MATBG by exploring the quantum metric length underlying the TBMs of the system, which could be responsible for the transport phenomena discovered in twisted bilayer graphene, e.g., the Josephson current found in~\cite{diez2025probing}. Concretely, we apply our theory on the 10-band lattice model for the twisted bilayer graphene introduced in \cite{PhysRevB.99.195455}. This example not only demonstrates the universal role of the QML, but also suggests a possible platform for detecting the quantum metric length.

In the chiral limit of the MATBG, there are two exact flat bands in the middle of the 10 bands, as plotted in Fig.~\ref{Fig_MATBG_band_QM_wavefunc_QS}(a), and these flat bands have the same origin as the BCL introduced in the main text.
%Eq.~(\ref{Eq_BCL_Hamilt}).
In Fig.~\ref{Fig_MATBG_band_QM_wavefunc_QS}(b), a logarithmic plot of the BZ distribution of the traced non-Abelian quantum metric summing over the band index pronouncedly exhibits a hot spot region around the $\Gamma$ point, where the band gap of the flat bands is located. Examining the Hamiltonian structure, we can see that it has a typical structure of the bipartite crystalline lattices (BCL). Interestingly, the off-diagonal block $S_{\bm{k}}$ can be partitioned into two parts, $S_{1, \bm{k}}$ and $S_{2, \bm{k}}$, with contrasting energy scales:
\begin{equation}
    S_{1, \bm{k}} = J\begin{pmatrix}
        -(\omega+\phi_{11}\omega^*+\phi_{01})\zeta^*a &-i\omega d & -i\phi_{01}\omega^*d & -i\phi_{\bar{1}0}d \\
        (1+\phi_{11}\omega^*+\phi_{01}\omega)\zeta^*a &-i\omega^* d & -i\phi_{\bar{1}0}d & -i\phi_{01}\omega d \\
        (\omega+\phi_{10}\omega^*+\phi_{11})\zeta a &i \omega d & i \omega^* d & -i d \\
        -(\omega+\phi_{10}+\phi_{11}\omega^*)\zeta a & i\omega^* d & i\omega d & i d \\
    \end{pmatrix}
    \label{Eq_TBG_S1}
\end{equation}
\begin{equation}
    S_{2, \bm{k}} = \kappa J\begin{pmatrix}
        -(\omega^*+\phi_{11}\omega+\phi_{01})\zeta b & (1+\phi_{11}+\phi_{01}) c \\
        (1+\phi_{11}+\phi_{01}) c & (1+\phi_{11}\omega+\phi_{01}\omega^*)\zeta b \\
        (\omega^*+\phi_{10}\omega+\phi_{11})\zeta^* b & (1+\phi_{10}+\phi_{11})c \\
        (1+\phi_{10}+\phi_{11}) c & (\omega^*+\phi_{10}+\phi_{11}\omega)\zeta^* b \\
    \end{pmatrix},
    \label{Eq_TBG_S2}
\end{equation}
where $\omega = e^{i{2\pi}/{3}}\zeta$, $\zeta^2 = \omega$, and $\phi_{lm} \equiv e^{-i\bm{k}\cdot(l\bm{a}_1+m\bm{a}_2)}$, with $\bm{a}_1$ and $\bm{a}_2$ being the crystal lattice vectors and $\bar{l}\equiv -l$. To reproduce the MATBG band structure, the parameters are taken as $J=130$meV, $a=0.11$, $b=c=0.033$, $d=0.573$, $\kappa = 1$. After partitioning, the block $S_{2,\bm{k}}$ has subdominant energy scales compared to $S_{1,\bm{k}}$, allowing us to implement our framework. 
Since the energy scale $\text{max}(|S_{2, \bm{k}}|)$ controls the gap between the flat bands and the higher bands, the non-Abelian quantum metric of the flat bands at the $\Gamma$ point could be enhanced by decreasing the energy scale of $S_{2, \bm{k}}$ using the parameter $\kappa$ ($0<\kappa \le 1$). In Fig.~\ref{Fig_MATBG_band_QM_wavefunc_QS}(f), we calculated the \(y\)-directional quantum metric length $\xi_{QM,y,k_x}$ near the $\Gamma$ channel ($\tilde{\bm{k}}=k_x=0$) with varying $\kappa$ using Eq.~(\ref{Eq_QML_definition_SM}). As expected, the $\Gamma$ channel QML is significantly enhanced when decreasing $\kappa$, while the QML for the channels away from $\Gamma$ has an opposite effect from $\kappa$. 

To make the flat bands topological, we introduce the coupling matrix $H_d^{(g)}$ with $d=2, g=1$ instead of the original coupling scheme in~\cite{PhysRevB.99.195455} for two reasons. First, it allows flexible control of the conventional behavior, and the geometric behavior can be exhibited solely when exploring the role of the QML. Secondly, the definition of the QML in Eq.~(\ref{Eq_QML_definition_SM}) only involves the degenerate exact flat bands, making our primary conclusion independent of how specifically topology is introduced. Therefore, the topological flat bands employed here provide a streamlined counterpart to those in the realistic MATBG, while retaining the essential physics when inspecting the geometric behavior and the role of the QML in TBMs.

Let us examine the role of the QML for the TBMs in this system. Imposing an OBC in the \(y\)-direction, we obtain its spectrum in Fig.~\ref{Fig_MATBG_band_QM_wavefunc_QS}(c), where two flat bands are highlighted in cyan. A zoomed-in view of the topological flat bands, along with the in-gap edge spectrum, is presented in Fig.~\ref{Fig_MATBG_band_QM_wavefunc_QS}(d). Here, the bulk states around the $\Gamma$ channel merge into the gap due to the comparable energy scales of flat bands' bandwidth and $S_{2, \bm{k}}$. It is also noticeable that, besides the highlighted ones, other in-gap boundary modes also emerge under OBC but do not possess a topological origin. Those TBMs exhibit two phases of spatial behavior consistent with our discussions in the main text: A conventional oscillatory decay originating from the flat bands dispersion, followed by an exponential decay controlled by the QML. To elucidate the role of the QML, we focus on boundary modes that exhibit purely geometric behavior with $\xi_c=0$ and examine their spatial spread $\Omega_{y,k_x}$. Nevertheless, not all channels can be well controlled to solely exhibit the geometric behavior; e.g., the $\Gamma$ channel has a very small gap, as explained earlier, making the flat-bands-determined conventional length a significant contribution to the spatial spread of the TBMs. Consequently, the $\Gamma$ channel TBMs have a considerably large spatial spread, far exceeding the QML, as shown in Fig.~\ref{Fig_MATBG_band_QM_wavefunc_QS}(g). Conversely, away from the $\Gamma$ point where the conventional length is suppressed, the spatial spread has the same dependence on $\kappa$ as the QML, suggesting a quantum-metric origin of the TBMs behavior in this regime. The ratio between the spatial spread and the QML is presented in Fig.~\ref{Fig_MATBG_band_QM_wavefunc_QS}(h), clearly indicating that the QML serves as a loose, yet stringent, lower bound for the spread of the TBMs in MATBG. Through this ten-band model, we highlight the ubiquitous presence the QML and its universal role in boundary-mode localization in topological flat-band systems.

%satisfying the conditions we proposed for realizing flat bands with tunable quantum metric in Eqs.(\ref{Eq_recover_low_energy_model_condition}) and (\ref{Eq_nontrivialize_FB_condition}).

%The main conclusion of this paper concerns the non-Abelian quantum metric of the degenerate flat bands, which is irrespective of how specifically the flat bands are coupled to capture the flat bands details, including the dispersion, symmetry and topology, etc. Therefore, we follow the Lieb-QWZ model to couple the flat bands as Eq.~(\ref{Eq_QWZ_hamilt}), enabling well controlling over the conventional behavior of TBMs when inspecting the geometric behavior and the role of QML.

\section{Supplementary Note IV: More discussions and applications of the QML}	

%\subsection{\label{appendix_another_TFB_model} Topological flat bands without involving the BCL}

%%%%%%%%%%%%%%%%%%%%%%%%%%%%%%%%%%%%%%%%%%%%%%%%%

\subsection{Alternative coupling schemes in constructing topological flat bands}
\label{appendix_coupling_schemes}
We have introduced the minimal topological coupling in the main text to generate topological flat bands by restricting the coupling term to act exclusively among the orbitals of $L_{2,B}$ in the partitioned BCL model as:
\begin{equation}
    H_{\text{c}}(\bm{k}) = \begin{pmatrix}
        O_{N_{L_1}\times N_{L_1}}  & O\\
        O  & H_{d}^{(g)}(\bm{k})
    \end{pmatrix},
\end{equation}
with $H_{d}^{(g)}(\bm{k}) = \sum_{i=1}^d 2t_i \sin{k_i} \gamma_i^{(g)} + \left(m - \sum_{i=1}^d 2\tilde{t}_i\cos{k_i}\right) \gamma_{d+1}^{(g)}$. 
For instance, when $d=1$, $g=1$ and $N_{L_1}=2$, the coupling matrix becomes:
\begin{equation}
    H_{\text{c},1}(\bm{k}) = \begin{pmatrix}
        O_{2\times 2}  & O_{2\times 2} \\
        O_{2\times 2}  & 
        \begin{matrix} 
            m- 2\tilde{t}_1\cos{k_1}  &  2t_1 \sin{k_1} \\
            2t_1 \sin{k_1}  & 2\tilde{t}_1\cos{k_1}) 
        \end{matrix}
    \end{pmatrix}.
    \label{Eq_coupling_scheme1}
\end{equation}

\begin{figure}[htbp]
    \centering
    \includegraphics[width=0.8\linewidth]{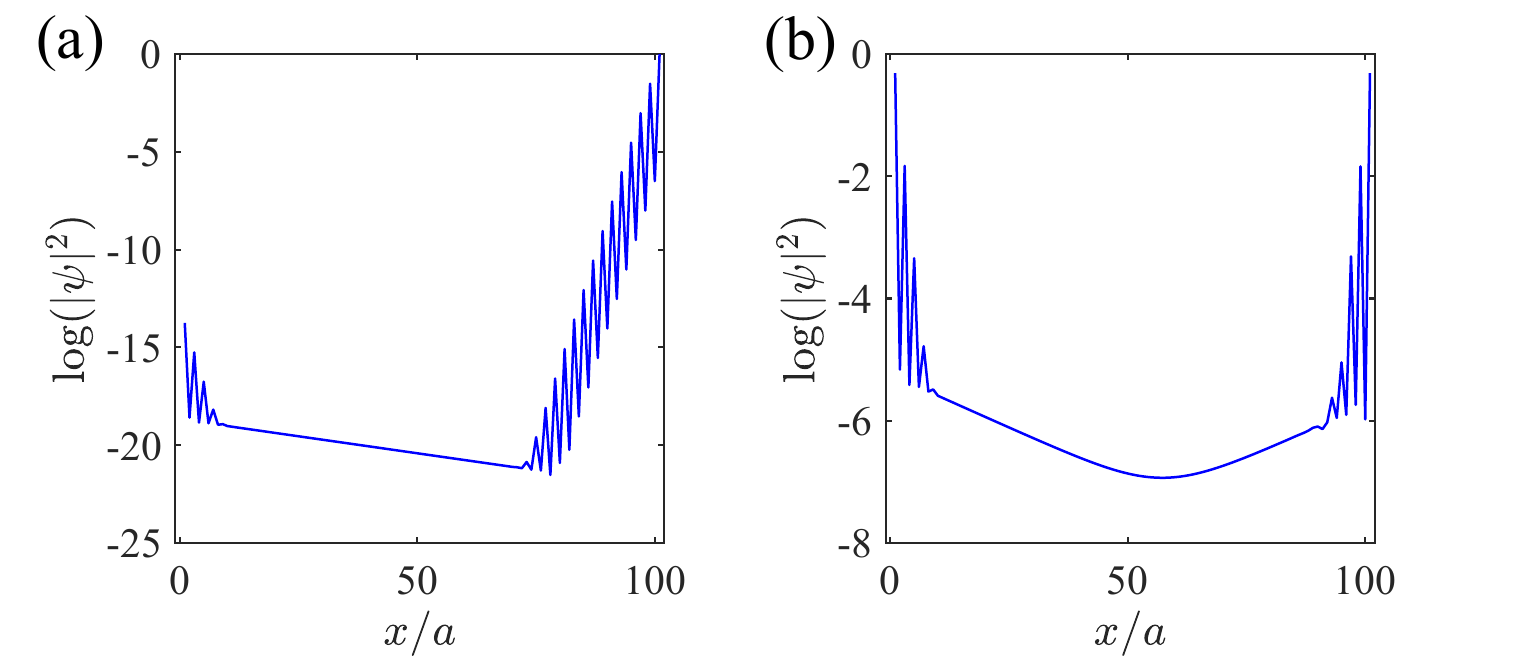}
    \caption{Logarithmic plots for the TBMs wave functions under the coupling scheme (a) $H_{\text{c},1}$ and (b) $H_{\text{c},2}$.}
    \label{Fig_Lieb_SSH_2coupling_schemes}
\end{figure}

Alternatively, we can introduce other coupling schemes on the degenerate manifolds to introduce topology by including more orbitals on the sublattice $L_2$. The following is a concrete example:
\begin{equation}
    H_{\text{c},2}(\bm{k}) = \begin{pmatrix}
        O_{1\times 1}  & O_{1\times 3} \\
        O_{3\times 1}  & 
        \begin{matrix} 
            0 & 2t_1 \sin{k_1} & 2t_1 \sin{k_1} \\
            2t_1 \sin{k_1} &  m- 2\tilde{t}_1\cos{k_1}  &  2t_1 \sin{k_1} \\
            2t_1 \sin{k_1} &	2t_1 \sin{k_1} &  -(m- 2\tilde{t}_1 \cos{k_1}) 
        \end{matrix}
    \end{pmatrix}.
    \label{Eq_coupling_scheme2}
\end{equation}
Both coupling schemes in Eqs.~(\ref{Eq_coupling_scheme1}) and (\ref{Eq_coupling_scheme2}) give rise to the same low-energy physics of the flat bands, thus preserving their topological properties. 
%However, different coupling schemes will fine tune how the boundary states are coupled by redistributing their orbital components, thereby influencing their transport effects, especially in 1D.
However, by including more orbitals in the coupling matrix such as Eq.~\eqref{Eq_coupling_scheme2}, the orbital components for later TBMs will be rearranged so that the boundary modes at opposite ends can have better overlaps, as presented in Fig.~\ref{Fig_Lieb_SSH_2coupling_schemes}.
The second coupling scheme is thus adopted in numerical calculations when exploring the resonance tunneling effect and the crossover in the flat-band QAH Fraunhofer patterns.

\begin{comment}
    
When further introducing coupling between orbital B and C(D) as:
\begin{equation}
	H_{\text{c},2}(\bm{k}) = \begin{pmatrix}
    		O_{1\times 1}  & O_{1\times 3} \\
    		O_{3\times 1}  & 
    		 \begin{matrix} 
        		 	0 & 2t_1 \sin{k_1} & 2t_1 \sin{k_1} \\
        		2t_1 \sin{k_1} &  m-\sum_{i=1}^2 2\tilde{t}_i\cos{k_i}  &  2t_1 \sin{k_1} - i 2t_2 \sin{k_2} \\
        		2t_1 \sin{k_1} &	2t_1 \sin{k_1}+ i 2t_2 \sin{k_2}  &  -(m-\sum_{i=1}^2 2\tilde{t}_i\cos{k_i}) 
        		\end{matrix}
    	\end{pmatrix},
	\label{Eq_coupling_scheme2}
\end{equation}

\end{comment}

%\emph{comment: As is well know, the non-trivial topology gives a finite lower bound to the quantum metric tensor. However, the spread of the boundary states is not lower bounded by the topological invariants like Chern number, winding number, etc. That's because the quantum metric bound is indeed summed over of all the topological bands, which ends up with zero. }

\subsection{\label{Sec_resonance_tunneling} Non-local resonance tunneling}
As pointed out in the main text, the TBMs generally possess two distinct and independent phases of behaviors: a conventional behavior resulting from bare band dispersion, and a geometric behavior controlled by the QML. The conventional length $\xi_c$ will be suppressed in the flat band limit, leaving only the geometric behavior dominating the real space physics. In this section, we will introduce some consequences resulting from this geometric behavior of TBMs tunable via the QML.
Transport measurements are one of the most effective experimental tools for investigating solid state physics. Here, we introduce a robust non-local resonance tunneling effect as the transport benchmark for detecting the QML in (quasi-)1D topological systems.

\begin{figure}[htbp]
    \centering
    \includegraphics[width=0.5\linewidth]{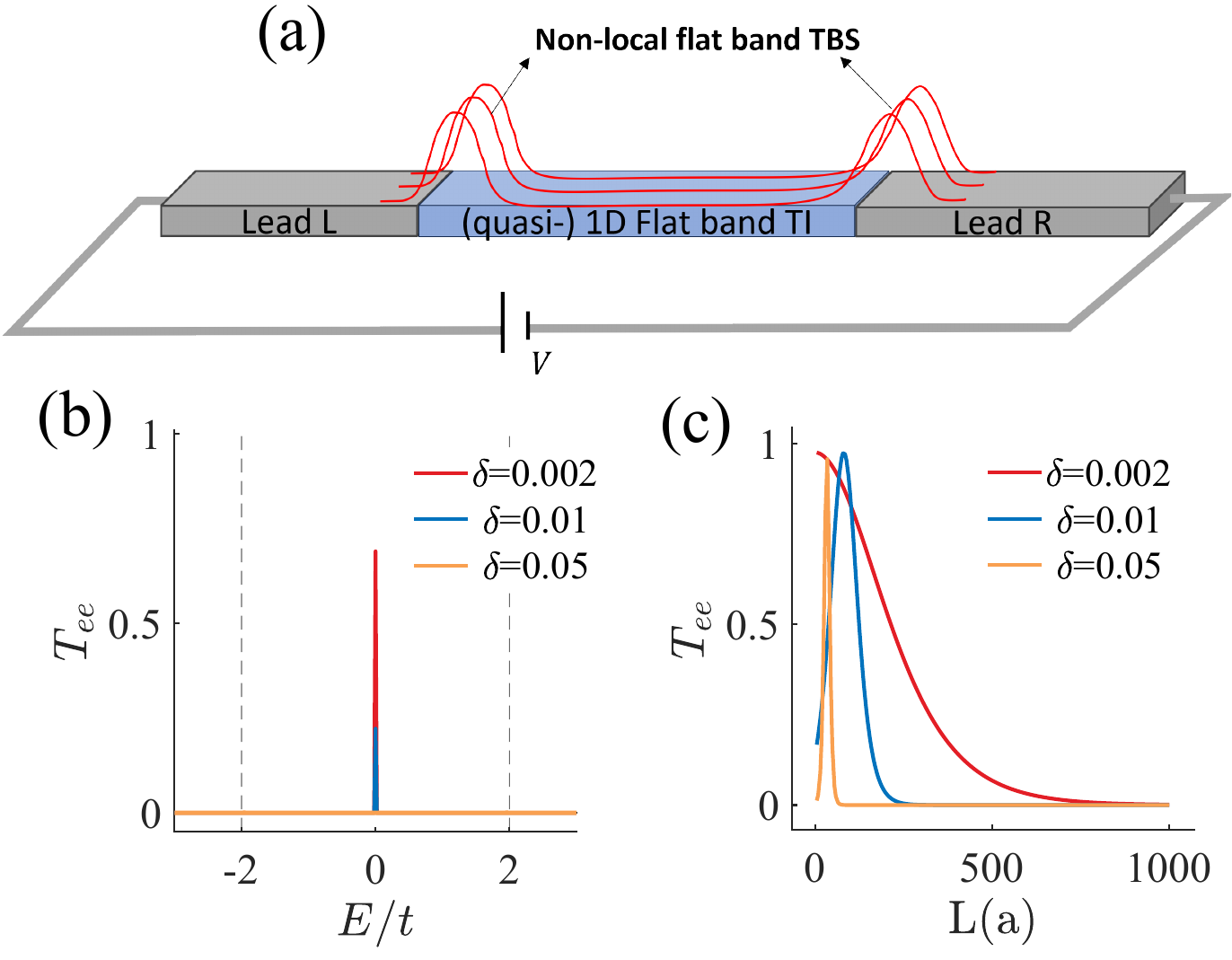}
    \caption{(a) An illustration of the resonance tunneling in a flat-band TI. The red lines represent the non-local flat-band topological boundary modes that are coupled due to a large QML. (b) The tunneling conductance carried by the non-local boundary modes under a bias voltage $E=eV$. The two dashed lines represent the topological gap, inside which is a tunneling peak. The length for the device is fixed at $L=150a$, the hopping strength inside the lead is  $t_l = \tilde{t}$, and the coupling between the lead and device are $t_{ld} = 0.02\tilde{t}$, respectively. Other parameters are $t = \tilde{t}$, $J=1$, $\tilde{t}= 2\times10^{-4}J$, $m = 2\tilde{t}$, $\delta = 0.05$ and $\alpha = 0.1$. (c) The tunneling conductance versus the device width $L$.}
    \label{Fig_Flatband_TI_tunneling}
\end{figure}

TBMs across the bulk can couple with each other when they are overlapped in real space, which enables non-local resonance tunneling inside the topological gap. With the QML this effect can be rather prominent. Specifically, consider a (quasi-) 1D TI hosting topological flat bands that is contacted by two metallic leads at its opposite ends, as illustrated in Fig.~\ref{Fig_Flatband_TI_tunneling}(a). When an electric volt is applied between the leads, one would usually expect a tunneling current to only appear outside the insulating gap, even with the presence of in-gap boundary channels. These channels do not conduct because they typically extend a few nanometers into the bulk, while the device is usually on the scale of hundreds of nanometers to micrometers. Therefore, they will not carry electric current. However, in topological flat-band systems with large QML, the TBMs can spread so widely enabling tunneling current in rather long devices, easily reaching one thousand unit cells, which is about several hundred nanometers. Consequently, we expect that the electrons carried by these modes can tunnel through the TI, resulting in an in-gap conductance peak.
	
We simulate such an experimental setup numerically, with the 1D flat-band TI being the $k_y=0$ channel of the proposed Lieb-QWZ model, 
%Lieb-SSH model
and calculate the tunneling conductance using the recursive Green function method. The scattering matrix element from channel $n$ in lead $j$ to channel $m$ in lead $i$ at zero temperature is given by \cite{datta1997electronic},
\begin{equation}
    t_{ij}^{mn} = -\delta_{ij}\delta_{mn} + i (\sqrt{\Gamma})^m_i (G^r)^{ij}_{mn}(\sqrt{\Gamma})^n_j.
    \label{Eq_scattering_matrix}
\end{equation}
Here, $G^r$ is the retarded Green's function. The broadening function $\Gamma$ relates to the imaginary part of the retarded self-energy $\Sigma^r$ by $\Gamma = -2\Im{(\Sigma^r)}$. In the setup of Fig.~\ref{Fig_Flatband_TI_tunneling}(a), $i/j=L/R$ denotes the left/right lead and $m,n$ are both the e-channels. Therefore, the electron tunneling probability will be $T_{ee} = |t^{ee}_{LR}|^2$. As shown in Fig.~\ref{Fig_Flatband_TI_tunneling}(b), inside the topological gap denoted as the region between the dashed lines, a prominent zero-bias peak for $T_{ee}$ arises. When decreasing the parameter $\delta$ with increased quantum metric length, the tunneling conductance is enhanced due to a stronger coupling between the boundary modes. On the other hand, a larger QML enables the tunneling process to persist in wider devices, as depicted in Fig.~\ref{Fig_Flatband_TI_tunneling}(c). The phenomenal ultra-long range resonance tunneling effect can be viewed as a benchmark for the large QML of the 1D topological flat-band  system.

\bibliography{SM_QMLTopologyNotes}

%\end{document}

		\end{document}